\documentclass[acmlarge]{acmart}

\makeatletter
\def\runningfoot{\def\@runningfoot{}}
\def\firstfoot{\def\@firstfoot{}}
\makeatother

\AtBeginDocument{%
  }

\usepackage{tikz}
\usetikzlibrary{trees}
\newtheorem{defn}{\textbf{Definition}}
\usepackage{tabularx}
\usepackage{multirow}
\usepackage{makecell}
\usepackage{threeparttable}
\usepackage{graphicx}
\usepackage{subfigure}
\usepackage{url}
\usepackage{enumitem}
\usepackage{cleveref}
\usepackage{wrapfig}

\begin{document}

\title{Machine Unlearning: A Comprehensive Survey}

\author{Weiqi Wang}
\email{Weiqi.Wang@ieee.org}
\affiliation{%
  \institution{University of Technology Sydney}
  \country{Australia}
}

\author{Zhiyi Tian}
\affiliation{%
	\institution{University of Technology Sydney}
	\country{Australia}}
\email{zhiyi.tian@student.uts.edu.au}

\author{Chenhan Zhang}
\affiliation{%
	\institution{University of Technology Sydney}
	\country{Australia}}
\email{chenhan.zhang@student.uts.edu.au}

\author{Shui Yu}
\affiliation{%
  \institution{University of Technology Sydney}
  \country{Australia}}
\email{shui.yu@uts.edu.au}

\renewcommand{\shortauthors}{Weiqi Wang et al.}


\begin{abstract}
	
	
As the right to be forgotten has been legislated worldwide, numerous studies have sought to design unlearning methods to protect users' privacy when they want to remove their data from machine learning service platforms. This has given rise to the concept of ``machine unlearning'' --- a field dedicated to removing the influence of specified samples from trained models. This survey aims to systematically classify a wide range of machine unlearning studies, discussing their differences, connections, and open problems. We categorize current unlearning studies into four key areas: traditional unlearning methods, unlearning verification, domain-centric machine unlearning, and privacy and security issues in unlearning. In the traditional unlearning part, we first classify traditional unlearning into exact unlearning and approximate unlearning at a higher level; then, we provide detailed introductions to the techniques used in these studies. Next, we review studies on unlearning verification to introduce how to assess the effectiveness of unlearning operations. Following the discussion of traditional unlearning and verification methods, we introduce domain-centric machine unlearning, including graph unlearning, federated unlearning, diffusion model unlearning, and large language model unlearning. Additionally, we consider privacy and security issues essential in machine unlearning and compile related literature. Finally, we discuss the challenges of various unlearning scenarios and highlight potential research directions.

\end{abstract}



\maketitle

\section{Introduction}
\label{sec:introduction}


Over the past decade, enormously increased data and fast hardware improvement have driven machine learning developments quickly.
Machine learning (ML) algorithms and artificial intelligence (AI) are embedded into day-to-day applications and wearable devices \cite{mahdavinejad2018machine}. It continuously collects increasing amounts of user information, including private data such as driving trajectories, medical records, and online shopping histories \cite{kirkpatrick2017overcoming,rolnick2019experience}. On the one hand, such an enormous amount of data helps to further advance ML and AI development. On the other hand, however, it poses a threat to users' privacy and creates a significant need for robust data management to ensure information security and privacy in ML \cite{tian2022comprehensive}. 

Machine unlearning has drawn growing research attention as the recent legislation of the "Right to be Forgotten" in many countries. Notable instances include the GDPR in the European Union \cite{mantelero2013eu}, the PIPEDA privacy legislation in Canada \cite{canada2018}, and the California Consumer Privacy Act in the United States \cite{de2018guide}. According to these laws, companies must take reasonable measures to guarantee that personal data is deleted upon request. It indicates that individual users have the right to request companies to remove their private data, which was previously collected for ML model training. The deletion is not only erasing their data from a database, but it also needs to delete the influence of the specified samples from trained models. The process of data removal from models was conceptualized as machine unlearning \cite{cao2015towards}.
Specifically, suppose a user (Alice) wants to exercise her right \cite{mantelero2013eu} when quitting a ML application, then the trained model of such application must "unlearn" her data. Such a process includes two steps: first, a subset of the dataset previously used for ML model training is requested to be deleted; second, the ML model provider erases the contribution of these data from the trained models. A naive data-erasing method is retraining a new ML model from scratch \cite{bourtoule2021machine}. However, the computation and storage costs of retraining are expensive, especially in complex learning tasks.

Many researchers have tried to find efficient and effective methods to implement unlearning rather than naive retraining, and there are several common challenges, which are summarized as follows. 
\textit{(1) Stochasticity of training}: A huge amount of randomness exists in the training process in machine learning, especially in complicated models' training periods such as CNNs \cite{gu2018recent} and DNNs \cite{sze2017efficient}. This randomness makes the training results non-deterministic~\cite{bourtoule2021machine} and raises challenges for machine unlearning to estimate the impact of the typical erased samples. 
\textit{(2) Incrementality of training}: The training process in machine learning is incremental, meaning that the model update from one data point influences the contribution of subsequent data points fed into the model. Deciding a way to effectively remove the contributions of the to-be-erased samples from the trained model is challenging for machine unlearning \cite{koh2017understanding}. 
\textit{(3) Catastrophe of unlearning}: 
\citeauthor{nguyen2020variational} \cite{nguyen2020variational} indicated that an unlearned model typically has worse model utility than the model retrained from scratch. The degradation would be {severe}, especially when a method tries to delete a huge amount of data samples. They referred to such sharp degradation as catastrophic unlearning \cite{nguyen2020variational}. 
Although several studies mitigate model utility degradation by bounding the loss function or restricting the unlearning update threshold, eliminating catastrophic unlearning remains an open problem. Recently, many studies have put efforts into solving these three main challenges and proposed many novel mechanisms that promote the progress of machine unlearning.

\begin{figure}[t]
	
	\Description{ARC}
	\tikzstyle{every node}=[draw=black,thick,anchor=west]
	\tikzstyle{selected}=[draw=black,fill=black!20]
	\tikzstyle{optional}=[draw=black,fill=black!40]
	\tikzstyle{bag} = [align=center]
	
	\begin{tikzpicture}[%
		grow via three points={one child at (-0.15,-0.7) and
			two children at (-0.15,-0.7) and (-0.15,-1.4)},
		edge from parent path={(\tikzparentnode.south)+(-.4cm,0pt) |- (\tikzchildnode.west)},
		]
		\node [draw=black,fill=black!40]{Machine Unlearning}
		child { node [selected] {Traditional Unlearning Methods}
				child { node {Exact Unlearning: \cite{cao2015towards,bourtoule2021machine,golatkar2021mixed,graves2021amnesiac,garg2020formalizing,brophy2021machine,yanarcane2022unlearning,chen2022recommendation,schelter2023forget,koch2023no,hu2024eraser}}
						child { node {Split Unlearning: \cite{cao2015towards,bourtoule2021machine,golatkar2021mixed,graves2021amnesiac,garg2020formalizing,brophy2021machine,yanarcane2022unlearning,chen2022recommendation}}}
				}
					child [missing] {}   
				child { node {Approximate Unlearning: \cite{wu2020deltagrad,wu2022puma,guo2019certified,sekhari2021remember,neel2021descent, schelter2021hedgecut,ginart2019making,golatkar2020eternal,mehta2022deep,li2020online, warnecke2021machine,nguyen2020variational,nguyen2022markov, fu2022knowledge, khan2021knowledge,tarun2023fast,wang2023machine,chundawat2023zero,warnecke2024machine,liu2023muter,lin2023erm,cha2024learning}
					}
				child { node {Certified Data Removal: \cite{wu2020deltagrad,wu2022puma,guo2019certified,sekhari2021remember,neel2021descent, schelter2021hedgecut,ginart2019making,golatkar2020eternal,mehta2022deep,li2020online, warnecke2021machine}}}
				child { node {Bayesian-based Unlearning: \cite{nguyen2020variational,nguyen2022markov, fu2022knowledge}}}
				} 
		}
		child [missing] {}           
		child [missing] {} 
		child [missing] {}  
		child [missing] {} 
		child [missing] {} 
		child { node [selected] {Unlearning Verification: \cite{hu2022membership, guo2023verifying, thudi2022unrolling, ma2022learn, sommer2022athena, huang2021mathsf, thudi2022necessity} } 
		}
			child { node [selected] {Domain-Centric Machine Unlearning}
						child { node  {Graph Unlearning: \cite{chien2022certified,chen2022graph,li2024towards,cong2022privacy,wang2023inductive,wu2023certified,zhang2024forgetting,wu2023gif}}} 
			child { node {Federated (Distributed) Unlearning Methods:  \cite{wang2022federated, wang2024forget, fraboni2024sifu, wang2023bfu, liu2021federaser, liu2022right, zhang2023fedrecovery, gupta2021adaptive, wu2022federated}}}
			child { node {Diffusion Model Unlearning: \citep{alberti2025data,chen2024score,wu2025erasing,li2024machine,zhang2024defensive,wu2025unlearning,cywinski2025saeuron,zhang2024unlearncanvas,ko2024boosting,george2025illusion}} }
			child { node {Large Language Models Unlearning: \citep{yao2024large,yao2024machine,liu2025rethinking,liu2024towards,huang2024offset,wang2025selective,liu2024large,jin2024rwku,li2024single}} }
		}
		child [missing] {} 
		child [missing] {} 
		child [missing] {} 
		child [missing] {}
		child { node [selected] {Privacy and Security Issues in Machine Unlearning} 
			child { node {Privacy Threats in Unlearning}
				child { node {Membership Inference: \cite{chen2021machine,zanella2020analyzing,shokri2017membership, lu2022label, huang2021mathsf,graves2021amnesiac,golatkar2020forgetting,Hu2024sp}}
				}
				child { node {Privacy Reconstruction: \cite{gao2022deletion,chen2021machine,zanella2020analyzing,shintre2019making,baumhauer2022machine,graves2021amnesiac,golatkar2020forgetting,Hu2024sp,zhang2023conditional}}
				}
			}  
			child [missing] {} 
			child [missing] {} 
			child { node {Security Threats in Unlearning : \cite{marchant2022hard,zhao2024static,di2022hidden,hu2023duty,liu2024backdoor}} }
			child { node {Unlearning Applications}
				child { node {Anomaly Removal: \cite{wang2019neural,wei2023shared,liu2022backdoor,cao2018efficient,du2019lifelong,Takashi2022learning, ye2022learning,ganhor2022unlearning}}}
				child { node {Data Unlearnable: \cite{huang2021unlearnable}}}  
			}
		};
	\end{tikzpicture}
	\caption{Our taxonomy for machine unlearning. The introduction order will also follow this figure. We classify the current unlearning literature into four main scenarios: traditional unlearning methods, unlearning verification, domain-centric machine unlearning, and privacy and security issues in machine unlearning. \vspace{-4mm}}
	\label{ARCt} 
\end{figure}
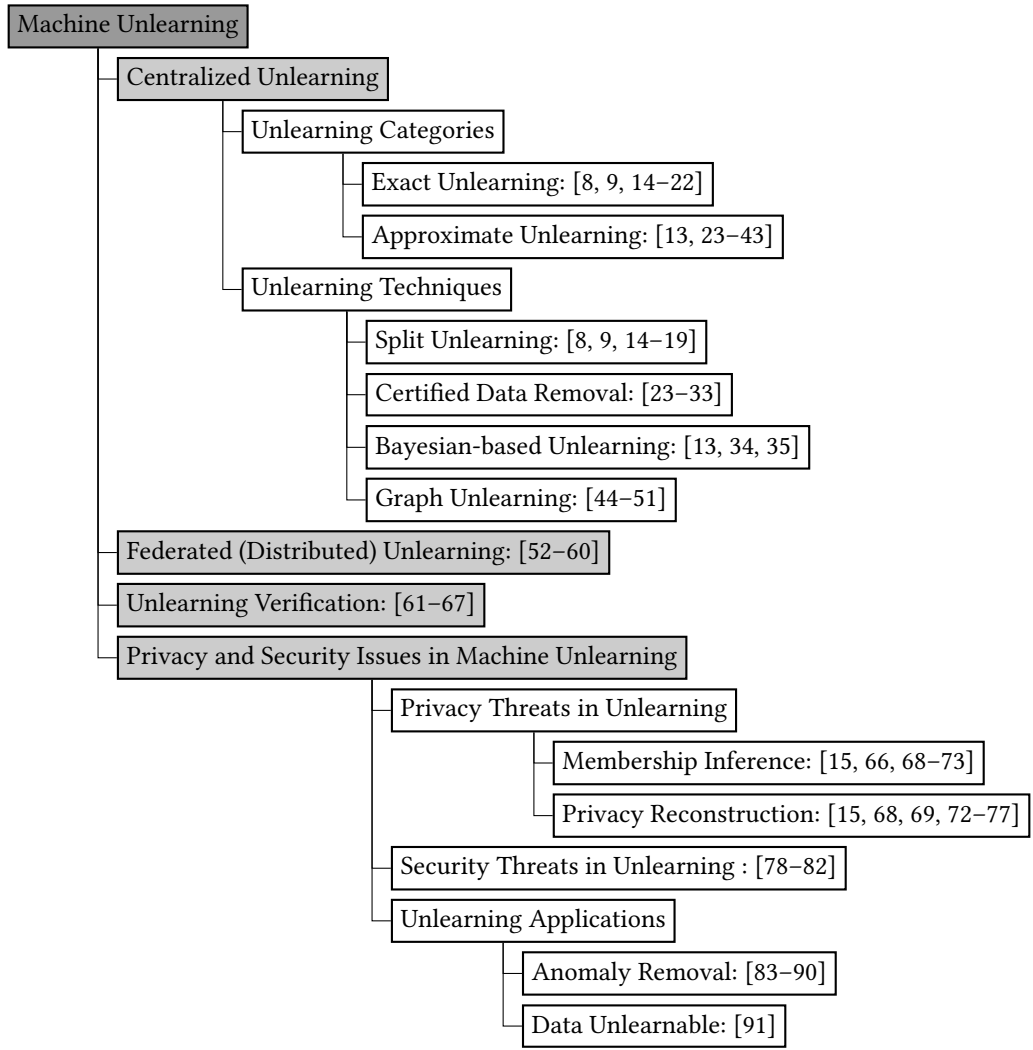

This survey aims to classify and systematize machine unlearning methods based on the research problems and objectives in the unlearning process and to review their differences, connections, as well as their advantages and disadvantages. {Our survey includes four main categories: traditional unlearning methods, unlearning verification, domain-centric unlearning methods, and privacy and security issues in machine unlearning, as shown in Fig. \ref{ARCt}. In the traditional unlearning category, we will introduce the representative \textit{Exact unlearning} \cite{cao2015towards,bourtoule2021machine} and \textit{Approximate unlearning} \cite{chien2022certified, nguyen2020variational} methods, including basic concepts and detailed technical implementation. After discussing traditional unlearning methods, verifying the unlearning effectiveness is equally crucial and has garnered significant research attention \cite{thudi2022necessity}. Therefore, we subsequently review related research on unlearning verification. Next, we will introduce domain-centric unlearning methods, including graph unlearning \citep{chien2022certified,chen2022graph}, federated unlearning~\citep{wang2022federated, liu2021federaser}, diffusion model unlearning \citep{alberti2025data,chen2024score}, and large language model (LLM) unlearning \citep{yao2024large,yao2024machine}.} Finally, we consider privacy and security issues in machine unlearning essential, organizing and reviewing the related publications on privacy and security threats, defenses, and unlearning applications.

 \begin{table*}[h]
 	\centering
 	\caption{ {Comparison of this survey with representative existing surveys on machine unlearning.}  \vspace{-2mm}
 	}
 	\label{survey_comparison}
 	\footnotesize
 	\setlength{\tabcolsep}{4pt}
 	\begin{tabularx}{\textwidth}{p{3.4cm}p{0.8cm}X X}
 		\toprule
 		\textbf{Survey} & \textbf{Year} & \textbf{Main focus} & \textbf{Difference from this survey} \\
 		\midrule
 		
 		\citeauthor{mahadevan2021certifiable}, \textit{Certifiable Machine Unlearning for Linear Models} \citep{mahadevan2021certifiable}
 		& 2021 
 		& Experimental review of approximate/certifiable unlearning methods for linear models, with emphasis on efficiency, effectiveness, and certifiability trade-offs. 
 		& We provide a broader taxonomy spanning traditional unlearning, verification, domain-centric unlearning, such as federated unlearning and LLM unlearning, and privacy/security issues. \\
 		
 		\citeauthor{nguyen2022survey}, \textit{A Survey of Machine Unlearning} \citep{nguyen2022survey}
 		& 2022 
 		& Broad overview of formulations, design criteria, removal requests, algorithms, scenarios, and applications. 
 		& Offers a strong general foundation, but places more emphasis on concepts and formulations than on systematically comparing scenario-specific challenges. \\
 		
 		\citeauthor{wang2023federated}, \textit{Federated Unlearning and Its Privacy Threats} \citep{wang2023federated}
 		& 2023 
 		& Federated unlearning, especially unlearner roles, privacy leakage, membership inference risks, and defenses. 
 		& Focused on federated settings only, particularly privacy threats in federated unlearning, rather than the full machine unlearning landscape. \\
 		
 		\citeauthor{liu2023survey}, \textit{A Survey on Federated Unlearning: Challenges, Methods, and Future Directions} \citep{liu2023survey}
 		& 2024 
 		& Dedicated taxonomy of federated unlearning methods, workflows, challenges, applications, and future directions. 
 		& Comprehensive for federated unlearning, but not designed to compare federated unlearning with centralized unlearning, verification, and broader privacy/security issues within one unified framework. \\
 		
 		\citeauthor{liu2025threats}, \textit{Threats, Attacks, and Defenses in Machine Unlearning: A Survey} \citep{liu2025threats}
 		& 2025 
 		& Security-oriented review of vulnerabilities, attacks, and defenses in machine unlearning systems. 
 		& Centers on threat taxonomy and defensive mechanisms, rather than on the full algorithmic and scenario-based landscape of machine unlearning. \\
 		
 		\textbf{This survey} 
 		& -- 
 		& \textbf{Unified, SLR-based survey covering traditional centralized unlearning, unlearning verification, domain-centric unlearning, privacy/security issues, applications, open questions, and their interconnections.} 
 		& \textbf{Distinctive in integrating previously separate strands into one taxonomy and explicitly comparing their differences, connections, advantages, limitations, and research gaps.} \\
 		
 		\bottomrule
 	\end{tabularx}
 \end{table*}

There are only a few surveys on machine unlearning because it is a relatively new research domain, and most of them are preprints on Arxiv, focusing on certain unlearning scenarios. For an introduction to machine unlearning, including discussions on exact and approximate unlearning problems and their solutions through recently proposed methods, see \cite{mercuri2022introduction}. For information on provable machine unlearning for linear models, including algorithm introductions and experimental analysis, refer to \cite{mahadevan2021certifiable}. For an overview of federated unlearning, see \cite{liu2023survey,wang2023federated}, and for graph unlearning, see \cite{said2023survey}. While \citeauthor{nguyen2022survey} \cite{nguyen2022survey} summarized the general unlearning framework and added the unlearning verification part to it, they focused primarily on introducing problem formulations and technical definitions. {For an security-oriented review of vulnerabilities in machine unlearning systems, see \citep{liu2025threats}. We provide a comparison difference between existing surveys in \Cref{survey_comparison}.}



Compared with existing surveys on machine unlearning, we performed a systematic literature review (SLR) following the guidelines presented by \cite{keele2007guidelines} and as exemplified in \cite{kubrak2022prescriptive}. The main contributions are as follows.


\begin{itemize}
	\item {We systematically catalog machine unlearning studies into traditional unlearning methods, unlearning verification, domain-centric unlearning, and privacy and security issues in machine unlearning.} 
	\item For the traditional centralized unlearning scenario, which draws the most attention, we divide related studies into exact and approximate unlearning and illustrate the connections, pros, and cons among these works.    
	\item We further explore the privacy and security threats that target machine unlearning and discuss the applications of unlearning in defending against traditional security and privacy issues.
	\item We discuss different challenges in various machine unlearning areas, from traditional centralized unlearning to federated unlearning and generative model unlearning, unlearning verification, and privacy and security issues in unlearning. We list the related open questions of each scenario and present potential research directions to solve them. 
\end{itemize}


The survey's remaining sections are arranged as follows. We show a systematic literature review model of the survey in \Cref{slr_of_the_model}. The basic knowledge of machine unlearning and related technical tools are summarized in Section \ref{pr}. {After introducing the background, we introduce the survey following the taxonomy order in Fig. \ref{ARCt}. The primary content of traditional unlearning is presented in Section \ref{Tax}, which includes two main unlearning categories and corresponding in-depth techniques. Section \ref{evaluation} introduces the unlearning evaluation and verification methods. We collect and introduce the unlearning methods for specific domains such as federated unlearning, graph unlearning, diffusion model unlearning, and LLMs unlearning, in Section \ref{new}. The privacy and security issues in machine unlearning are divided into two sections. Section \ref{priacy} introduces the privacy and security threats accompanied by machine unlearning. Section \ref{app} discusses machine unlearning applications, which are mainly applied in dealing with security issues. In Section \ref{discuss}, we discuss the challenges of current unlearning methods in a general way and enumerate the potential directions for further research.} At last, in Section \ref{conclu}, we provide a summary of the survey.

\section{The SLR Method of the Survey}  \label{slr_of_the_model}

We performed a systematic literature review (SLR) of the extant research on machine unlearning. The main goal was to compile a large body of research on machine unlearning and categorize them from various perspectives to facilitate analysis. Therefore, we employ the SLR method because it is especially appropriate for finding pertinent literature on a certain research topic \cite{keele2007guidelines}. The main process of the SLR method for the survey is presented as follows.


We first designed a search string according to the review protocol \cite{keele2007guidelines}, identified appropriate digital databases, and defined the data extraction strategy. We focused on the keywords ``unlearning'' and ``machine unlearning'' in the search string, formulated as \textit{``unlearning OR machine unlearning''}. We used this search string in IEEE Xplore, ACM Digital Library, Scopus, and the Web of Science to find relevant papers. Additionally, we conducted a search on Arxiv to identify further relevant literature. Consequently, a total of 972 papers were retrieved (the search was conducted on 7 July 2024). We then limited the publication years to those after 2020 and ensured that the keyword ``unlearning'' appeared in the title and abstract. Moreover, after filtering out duplicates and papers with fewer than six pages, 261 papers remained. Referring to the Google Scholar top publications and the China Computer Federation (CCF) recommendations lists, we focused on reviewing 103 papers from top venues. After including 33 references for related techniques, we reviewed a total of 136 references.

During the systematic literature review process, the data extraction strategy is defined as follows. Besides the basic information of papers (title, author, publication venue, year), we first categorize papers into the four main classes as Fig. \ref{ARCt} according to the research problems. Then, we extract the techniques utilized, the evaluating metrics, and the datasets employed in these papers, which will all be introduced in the following context.



\section{Background} \label{pr}

\subsection{Machine Unlearning}


There are several urgent demands driving machine unlearning research. The foremost is the demand for privacy preservation. With the ``right to be forgotten'' being legislated globally \cite{mantelero2013eu,canada2018}, machine unlearning ensures that users can request the erasure of their data from trained ML models. In addition to privacy concerns, other factors are promoting the development of machine unlearning. One significant factor is model utility. In the real world, vast amounts of data are generated daily, necessitating prompt updates to model services, as outdated data can negatively impact model performance \cite{wang2022learning,rolnick2019experience,lopez2017gradient}. An effective unlearning mechanism is crucial for mitigating the adverse effects of outdated and incorrect data on model utility. Another critical factor is security. Adversarial attacks and data poisoning \cite{tian2022comprehensive} can easily compromise deep learning models. Therefore, detecting and removing adversarial and poisoned data is essential to ensure model security. Once an attack type is identified, the model must erase the influence of these adversarial data using the unlearning mechanism.

\begin{figure}[t]
	\centering
	\includegraphics[width=1.\linewidth]{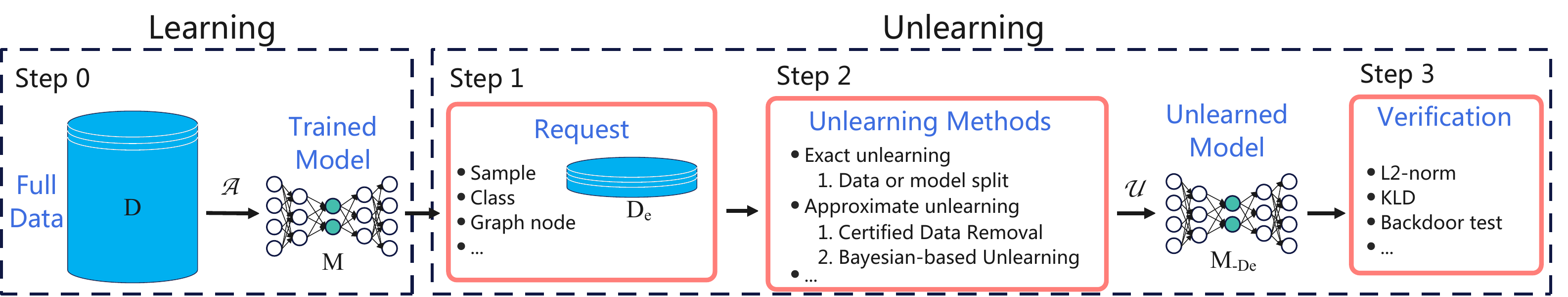}
	\caption{{Overview of the machine unlearning pipeline, from model training and unlearning request types to unlearning execution and verification methods.}
		\vspace{-2mm}}
	\label{fig:unlearningprocess}
\end{figure}

To better understand how unlearning mechanisms work, we first introduce the unlearning problem and process following the machine unlearning framework demonstrated in Fig. \ref{fig:unlearningprocess}, and we summarize the notations in \Cref{tab_symbols}.
Let $\mathcal{Z}$ denote a space of data items; the particular (full) training dataset is $D \in \mathcal{Z}$. A learning process can be demonstrated as Step 0 in Fig. \ref{fig:unlearningprocess}, i.e., training a model $M$ using an algorithm $\mathcal{A}$ on the training dataset $D$, denoted as $M = \mathcal{A}(D)$, where model $M$ is in a hypothesis space $\mathcal{H}$. 

The unlearning process begins at an unlearning request when a user wants to erase his specified data $D_e$ from the trained model, Step 1 in Fig. \ref{fig:unlearningprocess}. The requested unlearning data $D_e$ can be data samples \cite{bourtoule2021machine}, classes \cite{warnecke2021machine}, or graph nodes \cite{chen2022graph}. Then, in Step 2, the server removes the contribution of  $D_e$ using a designed machine unlearning algorithm $\mathcal{U}$. The unlearned model can be described as $M_{D \setminus D_e} = \mathcal{U}(M, D, D_e)$. {The standard aim of unlearning is to ensure the unlearned model $\mathcal{U}(M, D, D_e)$ is the same as the retrained model $\mathcal{A}(D  \setminus D_e)$ (i.e., $\mathcal{U}(M, D, D_e) \simeq \mathcal{A}(D  \setminus D_e)$).} Most unlearning studies ended at this step. However, we do need an effective evaluation metric to assess the unlearning effectiveness, as in Step 3 in Fig. \ref{fig:unlearningprocess}. Therefore, we add the verification step as the last step in the unlearning process, and we will introduce the related literature later.

\begin{table}[t]
	\caption{Notations in Machine Unlearning \vspace{-2mm}}
	\label{tab_symbols}
	\resizebox{0.9\linewidth}{!}{
		\begin{tabular}{cc|cc}
			\toprule
			Symbols & Description & Symbols & Description \\
			\midrule
			$\mathcal{Z}$ & data items space &  $\mathcal{H}$ & model parameters space \\
			$D = (X,Y)$& the training dataset with inputs $X$ and labels $Y$ &  $\mathcal{A}(\cdot)$ & the ML algorithm \\
			$D_e = (X_e, Y_e)$ & the unlearned (erased) dataset  &  $\mathcal{U}(\cdot)$ & the unlearning algorithm \\
			$D_r = D \setminus D_e$ & the remaining dataset & $M$ & the model with parameters $\theta$ \\
			$D_{\text{probe}}$ & the probing data used in attacking &  $M_{D \backslash D_e}$ & the unlearned model \\			
			\bottomrule
	\end{tabular}}
\end{table}

\subsection{Machine Unlearning Evaluation Metrics}

To compare the two models before and after unlearning, we need to define an evaluating metric $d(\cdot)$ between $\mathcal{A}(D \setminus D_e)$ and $\mathcal{U}(M, D, D_e)$. To this end, we briefly introduce several common evaluation distance metrics:

\noindent
\textbf{$L_2$-Norm.}
In \cite{wu2020deltagrad}, the authors propose utilizing the Euclidean distance to evaluate the parameters of the retrained model and the unlearned model. Let $\theta$ represent the model parameters learned by the algorithm $\mathcal{A}(\cdot)$. The $L_2$-norm measures the distance between $\theta_{\mathcal{A}(D \setminus D_e)}$ and $\theta_{\mathcal{U}(M, D, D_e )}$, where $\theta_{\mathcal{A}(D \setminus D_e)}$ are the model parameters retrained from scratch, and $\theta_{\mathcal{U}(M, D, D_e )}$ are the model parameters resulting from the unlearning algorithm $\mathcal{U}(\cdot)$.


\noindent
\textbf{Kullback–Leibler divergence (KLD)}. 
KLD is commonly used to measure the divergence between two probability distributions, often assessing the distance between retrained and unlearned models. In Bayes-based or Markov chain Monte Carlo-based unlearning methods~\cite{nguyen2020variational}, researchers utilize KLD \cite{joyce2011kullback} to optimize approximate models, employing it to measure the distance between two probability distributions. Recent unlearning studies have also used KLD to estimate the unlearning effectiveness by comparing the distributions of retrained and unlearned models \cite{nguyen2020variational}.

\noindent
\textbf{Evaluation Metric based on Privacy Leakage.}
Since membership inference attacks \cite{chen2021machine} can decide whether a sample was utilized for training a model, recently, some works have leveraged this property to verify if unlearning mechanisms remove the specific data. Some studies \cite{hu2022membership} even proposed to backdoor the unlearning samples for initial model training and then attack the unlearned model. If the unlearned model is still backdoored, this proves that the unlearning algorithm cannot unlearn samples effectively. Conversely, if the backdoor trigger cannot attack the unlearned model, it proves that the unlearning algorithm is effective.
Similar methods were also used in \cite{thudi2022unrolling, ma2022learn} to evaluate the unlearning effectiveness.

\subsection{Tools Used in Unlearning}


\noindent
\textbf{Differential Privacy (DP).}
Differential privacy is a popular benchmark for privacy protection in the Statistic \cite{dwork2006calibrating}. In a DP model, a trusted analyzer collects users' raw data and then executes a private method to guarantee differential privacy. The DP protection ensures the indistinguishability for any two outputs of neighboring datasets, where neighboring datasets mean the dataset only differs by replacing one user's data, denoted as $X \backsimeq X'$. 
A ($\epsilon, \delta$)-differential privacy algorithm $\mathcal{M}:\mathbb{X}^{n} \to \mathbb{Z}$ means that for every neighboring dataset pair $X \backsimeq X' \in \mathbb{X}^n $ and every subset $S \subset \mathbb{Z}$ has that $\mathcal{M}(X) \in S$ and $\mathcal{M}(X') \in S$ are $\epsilon$-indistinguishable and $\delta$-approximate. The degree of privacy protection rises with decreasing $\epsilon$. When $\epsilon = 0$, it implies that the outputted probability distribution of mechanism $\mathcal{M}$ cannot represent any meaningful information.
A general DP mechanism based on adding Laplace noise was presented and theoretically analyzed in \cite{dwork2006calibrating}.

\noindent
\textbf{Bayesian Variational Inference.}
In machine learning, the Bayesian variational inference is used to approximate difficult-to-compute probability densities via optimization \cite{kingma2014auto,blei2017variational}.
We revisit the variational inference framework that learns approximate posterior model parameters $\theta$ using Bayesian Theory in this part.
Suppose a prior belief $p(\theta)$ of an unidentified model and a complete data trainset $D$, an approximate posterior belief $q(\theta | D) \sim p (\theta | D)$ can be optimized by minimizing the KLD \cite{joyce2011kullback}, $\text{KL}[q(\theta | D) || p(\theta | D)]$. KLD measures how one probability distribution $q(\theta | D)$ differs from another probability distribution $p(\theta | D)$.
However, it is intractable to compute the KLD exactly or minimize the KLD directly. Instead, the evidence lower bound (ELBO) \cite{kingma2014auto} was proposed to be maximized, which is equivalent to minimize KLD between the two probability distributions.
ELBO follows directly from $\log(p(D))$ subtracting $\text{KL}[q(\theta|D) || p(\theta | D)]$, where $\log(p(D))$ is independent of $q(\theta| D)$. The ELBO is a lower bound of $\log(p(D))$ as $\text{KL}[q(\theta|D)|| p(D|\theta)] \geq 0$. In general training situations, ELBO is maximized using stochastic gradient ascent (SGA) \cite{kingma2014auto}.
The primary process is approximating the expectation $\mathbb{E}_{q(\theta|D)}[\log(p(D|\theta)) + \log (p(\theta)/q(\theta|D))]$ with stochastic sampling in each iteration of SGA. 
We can use a simple distribution (e.g., the exponential family)to approximate computational ease posterior belief $q(\theta |D)$.




\begin{figure}[t]
	\centering
	\includegraphics[width=0.8\linewidth]{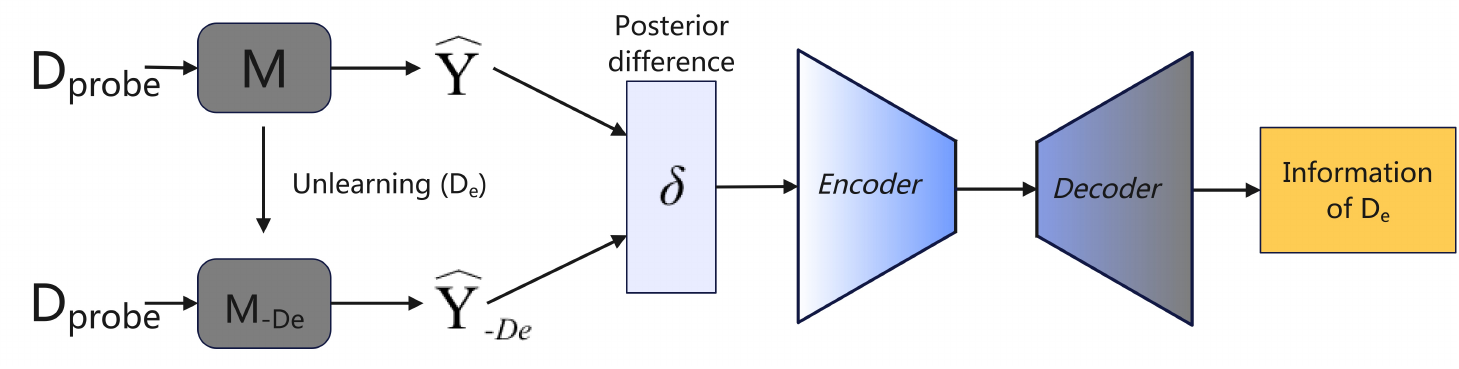}
	\caption{Privacy Leakage: a Privacy Reconstruction Process \vspace{-2mm}}
	\label{fig:attacksinunl}
\end{figure}


\noindent
\textbf{Privacy Leakage Attacks.} 
Privacy leakage occurs in both unlearning verification and privacy threats in two parts of unlearning. In unlearning verification, researchers tried to use privacy leakage attacks to verify whether the specific data is unlearned. Regarding the privacy and security issues in unlearning, researchers have tried to design effective inference attacks tailored to machine unlearning.
The basic attack of privacy leakage in a machine learning setting is membership inference, which determines if a sample was employed in the model updating process or not. When an attacker fully knows a sample, knowing which model was trained on it will leak information about the model. A generic membership inference process was introduced in \cite{shokri2017membership}. \citeauthor{shokri2017membership} first trained the shadow models to approach the target ML models. Then, they observed and stored the different outputs of the shadow models based on different inputs, in or not, in the trainset. They used these stored outputs as samples to train the membership inference attack model. 

Model inversion \cite{fredrikson2014privacy}, or privacy reconstruction \cite{salem2020updates} is another privacy threat in general machine learning. 
Model inversion aims to infer some lacking attributes of input features based on the interaction with the trained ML model. \citeauthor{salem2020updates} \cite{salem2020updates} proposed a reconstruction attack target recovering specific data samples used in the model updating by different model outputs before and after updating. Later, inferring the private information of updating data in conventional machine learning is transferred to inferring the privacy of the erased samples in machine unlearning.
In reconstruction attacks, the adversary first collects the different outputs using his probing data $D_{\text{probe}}$, including the original outputs $\hat{Y}_{M}$ before unlearning, and the outputs $\hat{Y}_{M_{-D_e}}$ after unlearning. Then, he constructs the attack model based on the posterior difference $\delta = \hat{Y}_{M_{-D_e}}- \hat{Y}_{M}$. The attack model contains an encoder and decoder, which has a similar structure as VAEs \cite{kingma2014auto}, and the main process is shown in Fig. \ref{fig:attacksinunl}.



\section{Traditional Centralized Machine Unlearning} \label{Tax}

In this section, we classify existing traditional centralized unlearning methods by their inherent mechanism and designed purposes and present the corresponding detailed techniques.


\subsection{Unlearning Solution Categories}

From the former introduction, we know that naive retraining is the most effective manner to realize machine unlearning. However, it is inefficient because it requires storing the entire original dataset and retraining the model from scratch, which consumes significant storage and computational resources, especially in deep learning scenarios. Therefore, researchers tried to design effective and efficient unlearning mechanisms, and two representative solutions are exact unlearning and approximate unlearning.


\subsubsection{Exact Unlearning} 


Exact unlearning is also called fast retraining, whose basic idea is derived from naive retraining from scratch.
Following the background of unlearning, we know the learning and unlearning algorithm, $\mathcal{A}(D)$ and $\mathcal{U}$, based on the trainset $D$ and erased dataset $D_e \subseteq D$, respectively. 
If $\mathcal{U}(\cdot)$ is implemented as naive retraining, the equality between $\mathcal{A}(D \setminus D_e) \in \mathcal{H} $ and $\mathcal{U}(M, D, D_e) \in \mathcal{H}$ is absolutely guaranteed.
However, naive retraining involves high computation and storage costs, especially for deep learning models and complex datasets \cite{thudi2022unrolling}. 
Unlike naive retraining, which relies on the whole remaining dataset, exact unlearning tries to retrain a sub-model only using a subset of the remaining dataset to reduce calculation cost.
A general operation of exact unlearning is that they first divide the dataset into several small sub-sets. Then, they transform the learning process by ensembling the sub-models trained with each sub-set as the final model \cite{bourtoule2021machine,cao2015towards}. 
So that when an unlearning request comes, they are just required to retrain the sub-model corresponding to the sub-set containing the erased data. They then ensemble the retrained sub-model and other sub-models as the unlearned model.

Exact unlearning aims to mitigate the computation cost when retraining a new model by transforming the original learning algorithms into an ensembling form. 
It divides the stochasticity and incrementality into several sub-models to reduce their influence. However, to some extent, they sacrificed the storage cost because they needed to store the whole training dataset in a divided form. 
\begin{itemize}[itemsep=0pt, parsep=0pt, leftmargin=*]
	\item In \cite{cao2015towards}, \citeauthor{cao2015towards} transformed the traditional ML algorithms into a summation form. They are only required to update several summations when an unlearning requirement comes, ensuring the method runs faster than retraining from scratch.
	\item SISA \cite{bourtoule2021machine} is a representative exact unlearning algorithm, which splits the full training dataset into shards and trained models separately in each shard. For unlearning, they simply need to retrain the shard that includes the erased data.  
	\item Study \cite{wu2022puma} proposed a framework that precisely models the impact of individual training sample on the model concerning various performance criteria and removes the impact of samples that are required to be removed. 
	\item \citeauthor{golatkar2021mixed} \cite{golatkar2021mixed} proposed an unlearning method on deep networks, splitting the trained model into two parts. The core part based on the data will not be deleted, and the unlearning part with the erased data will be unlearned with parameters bound. 
\end{itemize} 
These methods are efficient in computation, but they sacrifice the storage space to store the intermediate training parameters of different slices and the related training sub-sets.

Besides the high storage cost, another major issue with exact unlearning is that it is only suitable for scenarios where the unlearning request involves removing a few samples with low frequency.
Suppose an unlearning request needs to remove many data samples (usually, they are not in the same previous divided sub-set). In that case, exact unlearning must retrain all these related sub-models or even all the sub-models in the worst situation. At the moment, exact unlearning is no longer computation efficient, and the whole training dataset and intermediate parameters still need to be stored.

\subsubsection{Approximate Unlearning}

\begin{figure}
	\centering
	\includegraphics[width=0.9\linewidth]{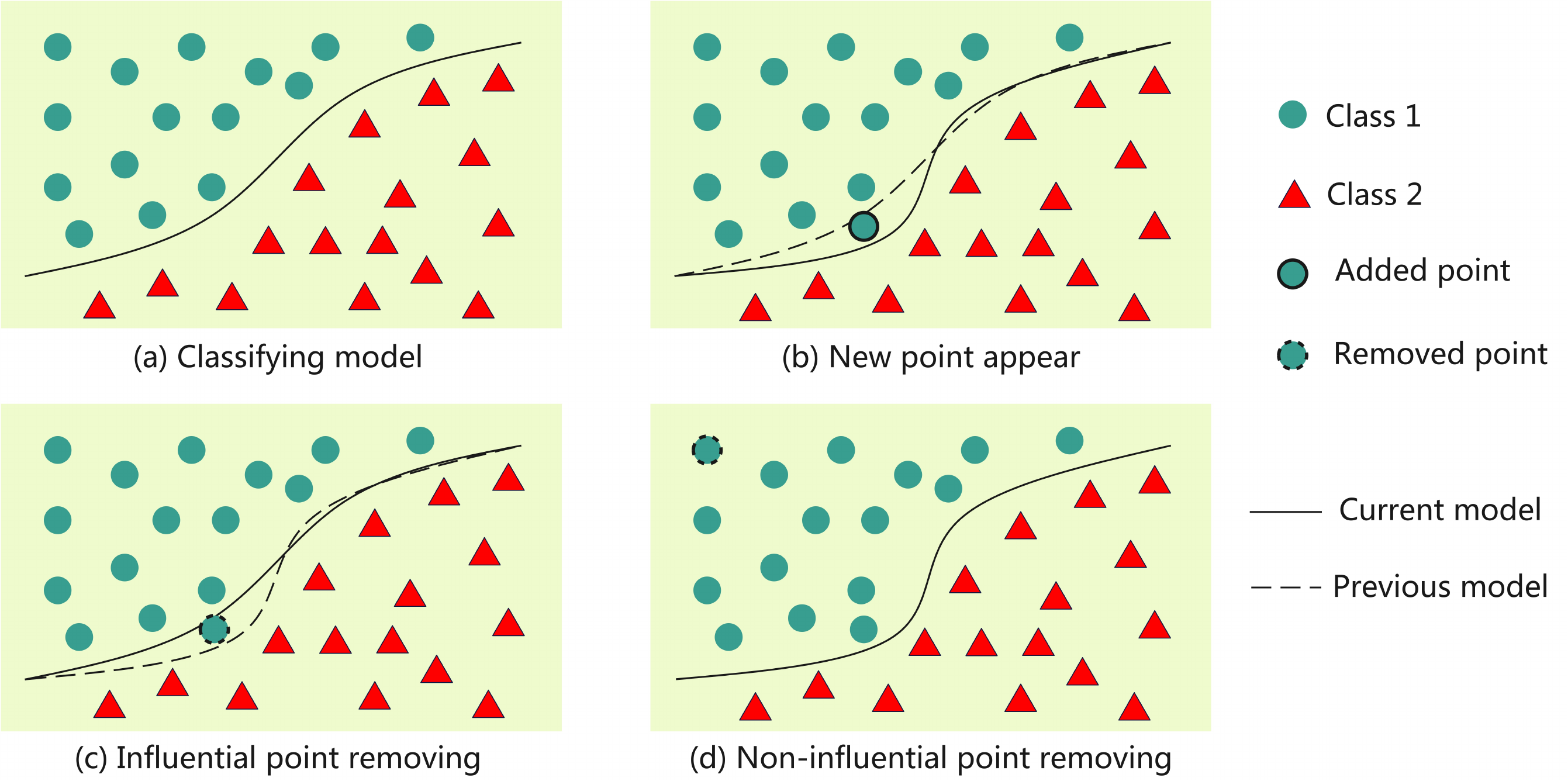} \vspace{-2mm}
	\caption{The model changes when adding a new point or removing a point. (a) A normally trained classifying model classifies classes 1 and 2. (b) When a new point appears, the model is trained based on it, and the classifying line is pushed to classify it. (c) When we need to remove an influential point, we should recover the contribution of this data point on the model. (d) When we remove a Non-influential point, the model may not need to change a lot. \vspace{-2mm}}
	\label{fig:logistic}
\end{figure}

Unlike exact unlearning, which only aims to reduce the retraining computation cost, approximate unlearning tries to directly unlearn based on the trained model and the erased data sample, which saves the computation and storage costs together.
Approximate unlearning studies aim to unlearn a model that approaches the model trained on the remaining dataset, i.e., the unlearned model $\mathcal{U}(M, D, D_e )$ should match the retrained model $\mathcal{A}(D \setminus D_e)$. Since exact unlearning is implemented by retraining from the remaining dataset or sub-sets, they can almost guarantee equality before and after unlearning.
However, since approximate unlearning tries to directly delete the influence of the unlearned samples from trained models, the core problem lies in precisely estimating and removing this contribution, which includes both stochasticity and incrementality.

The text description of the changes between two different distribution spaces before and after removing the specific data is not intuitive.
Fig. \ref{fig:logistic} shows illustrated changes when adding a new point or removing a point in a classifying model. When an influential point appears, it usually pushes the line to move forward than the original classifying line to identify it, as shown in Fig. \ref{fig:logistic} (b). When this influential point is requested to be removed, the unlearning mechanism must recover the model to the original one that has not been trained by this specific point, as shown in Fig. \ref{fig:logistic} (c).
However, when only unlearning a non-influential point, which may have almost non-influence on the model, the unlearned model may not change compared to the original trained model in this situation, as shown in Fig. \ref{fig:logistic} (d).

Many methods were proposed to implement approximate unlearning efficiently and effectively. The popular solutions are certified-removal \cite{guo2019certified} and Bayes-based mechanisms \cite{nguyen2020variational}, which are introduced in technical detail in Section \ref{Tech}. Although those techniques are approximately unlearning the contribution of all the erasing data, including the inputs and labels, they inevitably decrease the model accuracy to some extent after unlearning.

\noindent
\textbf{Main Challenges of Approximate Unlearning.} In centralized scenarios, researchers aiming at solving the basic machine unlearning problem will unavoidably face three challenges: stochasticity of training, incrementality of training, and catastrophe of unlearning. The exact unlearning methods extend the retraining idea, which avoids facing these challenges but consumes lots of storage costs. The approximate unlearning methods face these challenges directly, and we here list the relevant work about how to estimate the contribution of erased samples to overcome the stochasticity and incrementality of training, and how to prevent unlearning catastrophe.

\begin{itemize}
	\item To overcome the stochasticity and incrementality challenges when estimating the unlearning influence, one popular strategy is based on the first-order and second-order influence function \cite{basu2020second}, which is calculated based on the perturbation theory \cite{avrachenkov2013analytic}. {At the same time, since classical influence functions are best justified under smooth and strongly convex objectives, whereas deep neural networks usually involve highly non-convex loss landscapes, the local Taylor approximation and inverse-Hessian estimation can become unstable \cite{koh2017understanding,basu2020second}.} 
	\item The unlearning catastrophe appears commonly in approximate unlearning, and many studies try to propose some methods to solve this problem. In certified removal and Bayesian-based methods, they usually set a threshold to limit the unlearning update extent \cite{guo2019certified,nguyen2020variational}. In \cite{wang2023machine}, \citeauthor{wang2023machine} solves this problem by adding a model utility compensation task during unlearning optimization and finding the optimal balance based on multi-objective training methods.
\end{itemize}


\subsection{Detailed Techniques of Traditional Centralized Unlearning} \label{Tech}

\begin{table*}
	\Huge
	\caption{Traditional Unlearning Techniques \vspace{-2mm}}
	\label{tab:unl_tec}
	\resizebox{\linewidth}{!}{
		\begin{tabular}{cccccccc}
			\toprule
			\multirow{1}{*} {Unlearning Literature}  & \multirow{1}{*}{Taxonomy}       & \multirow{1}{*} {Requests Type} & \multirow{1}{*} {Techniques} & \multirow{1}{*} {Realization Method}  & \multirow{1}{*}{Year}\\
			\midrule
			SISA \cite{bourtoule2021machine}         & Exact unlearning                & Samples                         & Split Unlearning                & Data and Model Partition             & 2021 \\
			Amnesiac unl. \cite{graves2021amnesiac}  & Exact unlearning                & Samples                         & Split Unlearning                & Partially retraining                 & 2021 \\
			GraphEraser \cite{chen2022graph}         & Exact unlearning                & Graph nodes                     & Split Unlearning                & Data Partition                       & 2022 \\
			RecEraser \cite{chen2022recommendation}  & Exact unlearning                & Samples                         & Split Unlearning                & Balanced Data Partition              & 2022 \\
			ARCANE \cite{yanarcane2022unlearning}    & Exact unlearning                & Samples                         & Split Unlearning                & Parition by Class                    & 2022 \\
			HedgeCut \cite{schelter2021hedgecut}     & Exact unlearning                & Samples                         & Split Unlearning                & Tree ensemble learning               & 2021 \\
			DeltaGrad \cite{wu2020deltagrad}         & Exact unlearning                & Samples                         & Split Unlearning                & L-BFGS \cite{byrd1994representations}& 2020 \\
			ERASER \cite{hu2024eraser} &  Exact unlearning                & Samples                         & Split Unlearning  & Inference Serving-Aware & 2024 \\
			L-CODEC \cite{mehta2022deep}             & Approximate unlearning          & Samples                         & Certified Data Removal        & Markov Blanket selection             & 2022 \\
			PUMA \cite{wu2022puma}                   & Approximate unlearning          & Samples                         & Certified Data Removal        & SME                                  & 2022 \\
			Certified Removal \cite{guo2019certified}& Approximate unlearning          & Samples                         & Certified Data Removal        & LP                                   & 2019 \\
			$(\epsilon, \delta)$-unl. \cite{neel2021descent} & Approximate unlearning  & Samples                         & Certified Data Removal        & Perturbed gradient descent           & 2021 \\
			\cite{warnecke2021machine}                  & Approximate unlearning          & Samples                         & Certified Data Removal        & Influence Theory             & 2023 \\
			Graph unl. \cite{chien2022certified}     & Approximate unlearning          & Graph nodes                     & Certified Data Removal        & Certified removal                    & 2022 \\
			Gif \cite{wu2023gif}     & Approximate unlearning          & Graph nodes                     & Certified Data Removal        & Influence function                   & 2023 \\
			SUMMIT \cite{zhang2024forgetting}  & Approximate unlearning  	  & Graph nodes 	& Certified Data Removal  & Multi-Objective Optimization     & 2024 \\
			EUBO, rKL \cite{nguyen2020variational}   & Approximate unlearning          & Samples                         & Bayesian Unlearning     & VBI                                  & 2020 \\
			MCU \cite{nguyen2022markov}              & Approximate unlearning          & Samples                         & Bayesian Unlearning     & Monte cario-based                    & 2022 \\
			BIF \cite{fu2022knowledge}               & Approximate unlearning          & Samples                         & Bayesian Unlearning     & MCMC                                 & 2022 \\
			\bottomrule
	\end{tabular}}
	\begin{tablenotes}
		\footnotesize
		\item \textbf{Alogrihtm abbreviations}
		\item VBI: Variational Bayesian Inference, SME: Store medial estimation, MCU: Monte Cario-based machine unlearning, FIM: Fisher Information Matrix, MCMC: Markov chain Monte Carlo, BIF: Bayesian inference forgetting, LP: Loss perturbation, TF-IDF: Term Frequency Inverse Document Frequency, EUBO: Evidence upper bound, rKL: reverse Kullback–Leibler  
	\end{tablenotes}
\end{table*}

This section presents the popular techniques used in existing unlearning methods, in both exact and approximate unlearning studies. Exact unlearning extends the idea of retraining and tries to reduce the computation cost of unlearning. Approximate unlearning was proposed to find a way to reduce computation and storage consumption together. The dominant studies are summarized in Table \ref{tab:unl_tec}, where the primary technique used in exact unlearning is split learning. Two primary techniques used in approximate unlearning are certified data removal and Bayesian-based unlearning. 


\subsubsection{Split Unlearning}

\begin{figure}
	\centering
	\includegraphics[width=0.99\linewidth]{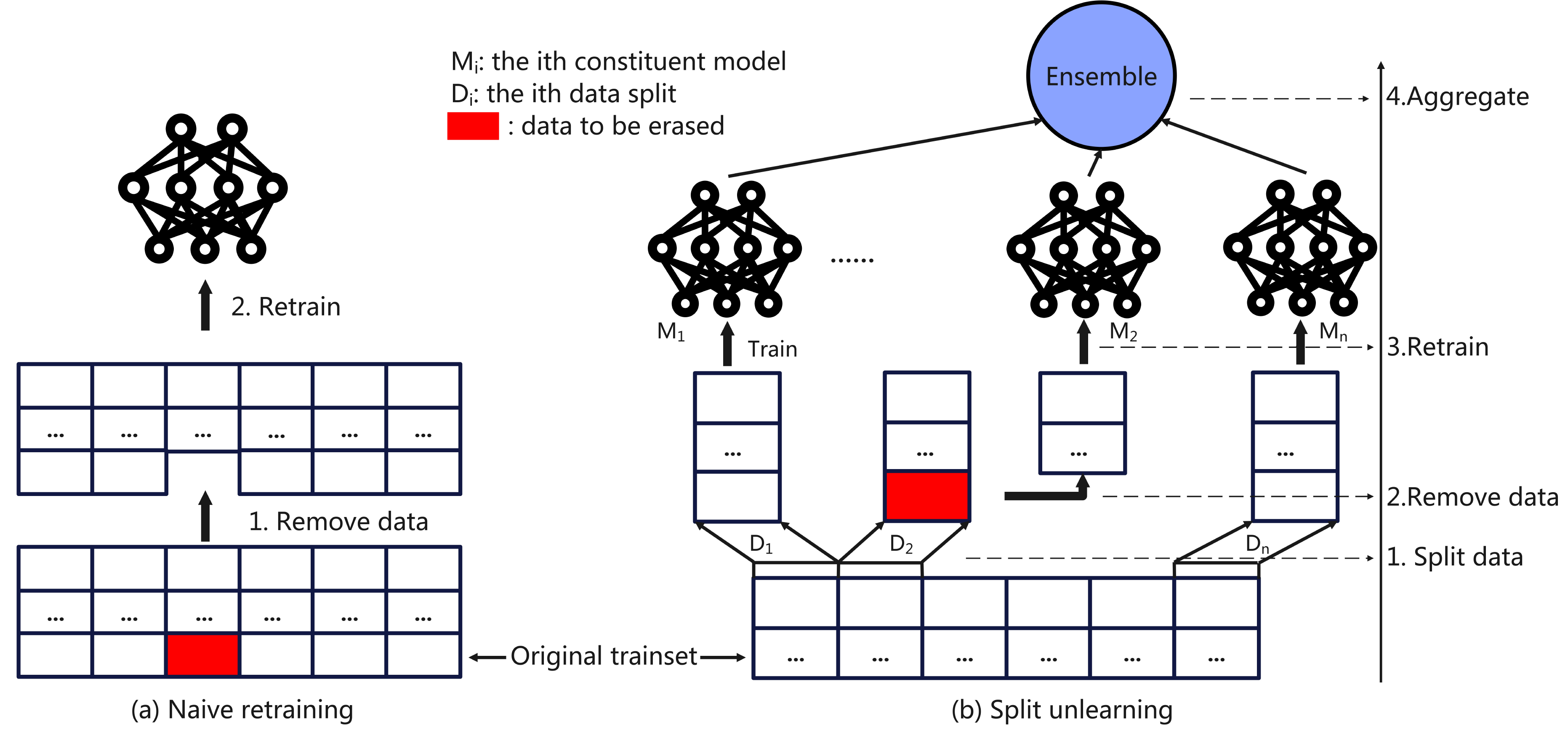}
	\caption{(a) Naive unlearning. There are only two steps: delete the specified samples from the whole dataset and retrain a model based on the remaining dataset. (b) Split unlearning. It contains four steps: 1. split the original dataset into $n$ shards, 2. remove the erased data from the corresponding shard, 3. retrain the sub-model of this shard, 4. ensemble all sub-models as the final model. \vspace{-4mm}}
	\label{fig:naiveandspliteunlearn}
\end{figure}

Since most exact unlearning methods attempted to partition the training dataset into multiple subsets and divide the ML model learning process, we call this kind of unlearning technique split unlearning. The main procedure of split unlearning is illustrated in Fig. \ref{fig:naiveandspliteunlearn}~(b). By contrast, the process of the naive retraining method is shown in Fig. \ref{fig:naiveandspliteunlearn}~(a), where we need two steps to realize it: first, delete the samples from the training dataset; second, retrain a new model using the remaining dataset. Since it needs to store the whole training dataset and retrain from the remaining dataset, it often entails significant computational and stored overhead. They proposed many exact unlearning methods to reduce the huge computation cost of naive retraining, and the majority of them are based on split learning techniques, either on data or model. As shown in Fig. \ref{fig:naiveandspliteunlearn} (b), the split unlearning technique can be summarized into four steps. Unlike naive unlearning trained based on the remaining trainset, the first phase of split unlearning is dividing the original full trainset into multiple disjoint shards. All the constituent models are trained based on each split data slice. Then, in the second phase, when the unlearning request comes, they only need to erase the requested samples from the split slice and retrain this slice's constituent model in the third phase. In the last phase, the split unlearning aggregates the retrained and other constituent models together as a new unlearned model.


The first split unlearning is proposed by \citeauthor{cao2015towards} \cite{cao2015towards}.
They split the original learning algorithms into a summation form. In a regular machine learning form, the model directly learns from the training dataset. However, in the summation form, they first train a small number of constituent models, which learn from several parts of the full trainset and then aggregate these intermediate models as the final learning model.
So that when unlearning, they only need to retrain the constituent model that contains the information of erased data. It can efficiently speed up retraining time and reduce computation costs.
In \cite{cao2015towards}, the authors indicated that support vector machines, naive Bayes classifiers, k-means clustering, and many ML algorithms could be implemented in a summation form to reduce the retraining cost. The statistical query (SQ) learning \cite{kearns1998efficient} guarantees the summation form. Although algorithms in the Probably Approximately Correct (PAC) setting can transform to the SQ learning setting, many complex models, such as DNNs, cannot be efficiently converted to SQ learning. 

Then, \citeauthor{bourtoule2021machine} \cite{bourtoule2021machine} and  \citeauthor{yanarcane2022unlearning} \cite{yanarcane2022unlearning} proposed advantaged methods unlearn samples suitable on deep neural networks. The primary idea of \cite{bourtoule2021machine, yanarcane2022unlearning} is also similar to the process shown in Fig. \ref{fig:naiveandspliteunlearn} (b). 
In \cite{bourtoule2021machine}, \citeauthor{bourtoule2021machine} named their unlearning method the SISA training approach. 
SISA can be implemented on deep neural networks, training multiple sub-neural networks based on divided sub-datasets. When the unlearning request comes, SISA retrains the model of the shard, which contains the information about the erased samples. SISA is effective and efficient as it aggregates all sub-models final prediction results rather than aggregates all these models. Unlike the original split unlearning dividing the dataset and transforming learning algorithms to summation form,
\citeauthor{yanarcane2022unlearning} proposed ARCANE \cite{yanarcane2022unlearning}, which transforms conventional ML into ensembling multiple one-class classification tasks.
When many unlearning requests come, it can reduce retraining costs, which was not considered in previous work.



\citeauthor{chen2022recommendation}~\cite{chen2022recommendation} extended exact unlearning methods to recommendation tasks and proposed RecEraser, which has similar architecture as split unlearning in Fig. \ref{fig:naiveandspliteunlearn} (b). RecEraser is tailored to recommendation systems, which can efficiently implement unlearning.
Specifically, they designed three data division schemes to partition recommendation data into balanced pieces and created an adaptive aggregation algorithm utilizing an attention mechanism.
They conducted the experiments on representative real-world datasets, which are usually employed to assess the effectiveness and efficiency of recommendation models.

Besides the above popular ML models, \citeauthor{schelter2021hedgecut} proposed HedgeCut \cite{schelter2021hedgecut}, which implemented machine unlearning on tree-based ML models in a split unlearning similar form.
Tree-based learning algorithms are developed by recursively partitioning the training dataset, locally optimizing a metric such as Gini gain \cite{wehenkel1991decision}. HedgeCut focuses on implementing fast retraining for these methods. Furthermore, they evaluated their method on five publicly available datasets on both accuracy and running time. 


Another method that is similar to split unlearning is Amnesiac Unlearning \cite{graves2021amnesiac}.
The intuitive idea of Amnesiac Unlearning is to store the parameters of training batches and then subtract them when unlearning requests appear.
In particular, it first trains the learning model by adding the total gradients $\sum_{e=1}^{E}\sum_{b=1}^{B}  \nabla _{\theta_{e,b}}$ to the initial model parameters $\theta_{\text{initial}}$, where $E$ is the training epochs, and $B$ is the data batches. In the model training process, they kept a list called $SB$, which records the batches holding the private data.
This list could be formed as an index of batches for each training example, an index of batches for each category or any other information expected.
When the unlearning request comes, a model using Amnesiac unlearning needs only to remove the updates from each batch $sb \in SB$ from the learned model $\theta_{M}$. As \citeauthor{graves2021amnesiac}~\cite{graves2021amnesiac} stated, using Amnesiac unlearning effectively and efficiently removes the contribution of the erased samples that could be detected through state-of-the-art privacy inference attacks and does not degrade the accuracy of the model in any other way.

\subsubsection{Certified Data Removal}

\begin{wrapfigure}{r}{0.5\textwidth}
	\centering
	\vspace{-4mm}
	\includegraphics[width=0.96\linewidth]{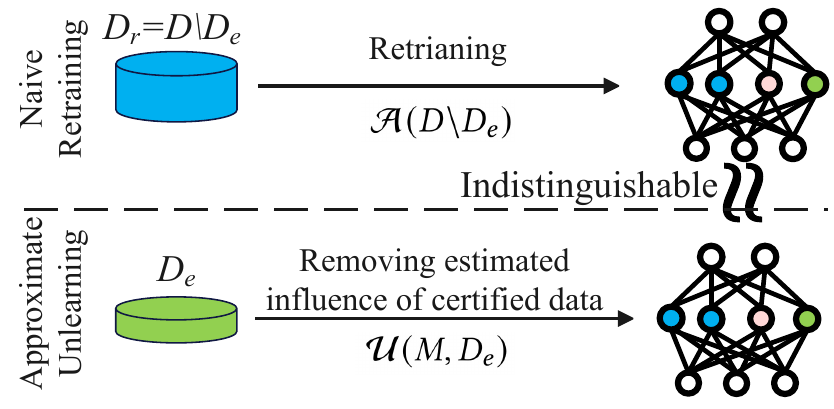}
	\vspace{-4mm}
	\caption{The approximate unlearning by certified data removal. The unlearning algorithm $\mathcal{U}$ includes estimating the influence of specified data and removing the estimation from trained models. The unlearned model is expected to approach the retrained model.}
	\label{fig_hessianbasedapproximate}
\end{wrapfigure}

Certified data removal unlearning methods usually define their unlearning algorithms as $\epsilon$-indistinguishable unlearning, which is similar to the differential privacy definition \cite{DworkR14}. An example is presented in \Cref{fig_hessianbasedapproximate}.
Most of them use the Hessian matrix \cite{bishop1992exact} to evaluate the contribution of erased data samples for unlearning subtraction. After estimating the impact of the erased data samples, they unlearn by subtracting these impacts with an updating bound from the unlearning model.

In \cite{guo2019certified}, \citeauthor{guo2019certified} proposed a certified data removal method, which assumes removing the last training sample, $(x_n, y_n)$. 
Specifically, they defined a removal mechanism that approximately minimizes $\mathcal{L}(\theta; D')$ with $D' = D \backslash (x_n, y_n)$. The loss gradient at sample $(x_n, y_n)$ can be denoted as $\Delta = \lambda \nabla \ell (\theta^{T} \cdot x_n, y_n)$ and the Hessian of $L( \cdot; D')$ at $\theta$ by $H_{\theta} =\nabla^2 L(\theta;D')$. 
Then, they applied a one-step Newton update to the model parameters impact of the erased point $(x_n, y_n)$ on the model $\theta$.
Under their observation, they found that directly removing the Hessian contribution from the gradient will reveal the private information of the erased data. They used the loss perturbation technique \cite{chaudhuri2011differentially} to hide this information. It used a random linear term to perturb the empirical risk and ensure that the outputs of their method $\mathcal{U}(D, D_e, \mathcal{A})$ is $\epsilon$-indistinguishable between the retrained model $\mathcal{A}(D\backslash D_e)$.
In\cite{mehta2022deep} and \cite{thudi2022unrolling}, they designed unlearning algorithms following the certified data removal definition in \cite{guo2019certified}.
\citeauthor{mehta2022deep}~\cite{mehta2022deep} unlearned via their proposed efficient Hessians, L-FOCI \cite{mehta2022deep}.
\citeauthor{thudi2022unrolling} \cite{thudi2022unrolling} used membership inference as a verification error to adjust the unlearning process on stochastic gradient descent (SGD) optimization.

Another similar unlearning method is PUMA. In \cite{wu2022puma}, \citeauthor{wu2022puma} proposed a new data removal method through gradient re-weighting called PUMA, which also used the Hessian Vector Product (HVP) term. They first estimated and recorded individual contributions of $(x_i, y_i)$, where the estimation is limited to less than one dot product between the pre-cached HVP term and individual gradient.
When the unlearning request comes, they subtract the estimate of the erased samples to revise the model.

\citeauthor{ginart2019making} \cite{ginart2019making} extended certified data removal to k-means clustering algorithms. They formulated the unlearning problem of efficiently removing personal data information from trained clustering models. They offered two different deletions for k-means clustering, quantized k-mean and divide-and-conquer k-means. In their work, both algorithms have theoretical guarantees and strong empirical results.  


To retrain SGD-based models fast, DeltaGrad was proposed by \citeauthor{wu2020deltagrad}~\cite{wu2020deltagrad} to unlearn small changes of data inspired by the idea of "differentiating the optimization path" concerning the training dataset and Quasi-Newton methods. 
They theoretically proved that their algorithm could approximate the right optimization path rapidly for the strongly convex objective.
DeltaGrad starts with a "burn-in" period of first iterations, where it computes the full gradients precisely. 
After that, it only calculates the complete gradients for every first iteration. For other iterating rounds, it operates the L-BGFS algorithm \cite{byrd1994representations} to compute Quasi-Hessians approximating the true Hessians, keeping a set of updates at some prior iterations. 

For a deeper understanding of certified machine unlearning, \citeauthor{sekhari2021remember} \cite{sekhari2021remember} further given a strict separation between $\epsilon$-indistinguishable unlearning and differential privacy. Different from \cite{guo2019certified}, in order to utilize tools of differential privacy (DP) for ML, the most straightforward manner is to forget the special dataset of erasure demands $D_e$ and create an unlearning mechanism $\mathcal{U}$ that solely relies on the learned algorithm $\mathcal{A}(D)$. In particular, the unlearning method is of the form $\mathcal{U}(D_e, \mathcal{A}(D)) = \mathcal{U}(\mathcal{A}(D))$ and makes sure the true unlearned model $\mathcal{U}(\mathcal{A}(D))$ is $\epsilon$-indistinguishable to $ \mathcal{U}(\mathcal{A}(D\backslash D_e))$. Notice the difference between \cite{guo2019certified} and \cite{sekhari2021remember}. In the definition of \cite{guo2019certified}, their $\epsilon$-indistinguishable unlearning is between $\mathcal{U}(\mathcal{A}(D))$ and $\mathcal{A}(D\backslash D_e)$, but here is between $\mathcal{U}(\mathcal{A}(D))$ and $\mathcal{U}(\mathcal{A}(D\backslash D_e))$.
Such a pair of algorithms in \cite{sekhari2021remember} would be differential private for $D$, where the neighboring datasets mean that for two datasets with an edit distance of $m$ samples.
The guarantee of DP unlearning is more powerful than the model distribution undistinguishable unlearning in \cite{guo2019certified}, and therefore, it suffices to satisfy it.
Based on the definition of \cite{sekhari2021remember}, they pointed out that any DP algorithm automatically unlearns any $m$ data samples if they are private for datasets with the distance $m$. Therefore, they derivate the bound on deletion capacity from the standard performance guarantees for DP learning. Furthermore, they determine that the existing unlearning algorithms can delete up to $\frac{n}{d^{1/4}}$ samples meanwhile still maintaining the performance guarantee w.r.t. the test loss, where $n$ is the size of the original trainset, and $d$ is the dimension of trainset inputs.

The aforementioned methods are trying to address the basic unlearning problem from a certified data removal perspective. \citeauthor{neel2021descent} \cite{neel2021descent} extended the definition of unlearning to include updates, which encompass both "add" ($D \cup \{z\}$) and "delete" ($D \backslash \{z\}$), where $z= (x,y)$ and $z \in \mathcal{Z}$ is a data point. They follow the definition in \cite{guo2019certified} and define similar $(\epsilon, \delta)$-indistinguishability. Furthermore, they extend $(\epsilon, \delta)$-indistinguishability to both "add" and "delete" updates, which can also be denoted as $(\epsilon, \delta)$-publishing. 

\subsubsection{Bayesian-based Unlearning }



Different from certified data removal that unlearns samples by subtracting corresponding Hessian matrix estimation from trained models, Bayesian-based unlearning tries to unlearn an approximate posterior as the model is trained by employing the remaining dataset. The exact Bayesian unlearning posterior can be derived from the Bayesian rule as $p(\theta | D_r) = p(\theta | D) \ p(D_e | D_r) / p(D_e | \theta)$, where $\theta$ is the posterior (i.e., model parameters). The erased dataset and the remaining dataset are two independent subsets of the full training dataset. If the model parameters $\theta$ are discrete-valued, $p(\theta|D_r)$ can be directly obtained from the Bayesian rule \cite{nguyen2020variational}. Additionally, employing a conjugate prior simplifies the unlearning process.

Nevertheless, it is challenging to get the exact posterior in practice, not to mention the unlearning posterior.
In \cite{nguyen2020variational}, \citeauthor{nguyen2020variational} 
tried doing likewise at the beginning.
They defined the loss function using the KLD between the approximate predictive distribution $q_u(y|D_r)$ and the exact predictive distribution $p(y|D_r)$. They bounded this loss function by the KLD between posterior beliefs $q_u(\theta | D_r)$ and $p(\theta | D_r)$ and further proposed evidence upper bound (EUBO) as the loss function to unlearn the approximate unlearning posterior.
To avoid the overestimation of using KL divergence to optimize the posterior, they introduced an adjusted likelihood to control the unlearning extent.

In \cite{nguyen2022markov}, the authors also studied the problem of "unlearning" particular erased subset samples from a trained model with better efficiency than retraining a new model from scratch. 
Toward this purpose, \citeauthor{nguyen2022markov}~\cite{nguyen2022markov} proposed an MCMC-based machine unlearning method deriving from the Bayesian rule. 
They experimentally proved that MCMC-based unlearning could effectively and efficiently unlearn the erased subsets of the whole training dataset from a prepared model.

\citeauthor{fu2022knowledge} \cite{fu2022knowledge} converted the MCMC unlearning problem into an explicit optimization problem. Then they proposed $\epsilon-$knowledge removal, which was a little similar to certified removal methods, but they defined that KLD between unlearned and retrained models must be less than $\epsilon$. To quantify the explicit $\epsilon-$knowledge removal, they proposed a knowledge removal estimator to assess the difference between the original and unlearned distributions. As they indicated, though their algorithm cannot wholly remove the learned knowledge from the already trained distribution, their method can still help the unlearned model approach its local minimum.

\section{Unlearning Evaluation and Verification} \label{evaluation}


Recent studies paid a huge amount of attention to unlearning problem-solving; however, verifying the unlearning effectiveness is also an important problem in machine unlearning. In this section, we will introduce some common and basic unlearning verification methods and evaluation datasets.

\begin{table*}[t]
	\Huge
	\caption{Evaluation and Verification Metrics \vspace{-2mm}}
	\label{tab_unl_verify}
	\resizebox{\linewidth}{!}{
		\setlength\tabcolsep{7.pt}
		\begin{tabular}{c|l|l|c}
			\toprule
			\multirow{1}{*} {Evaluation Metrics}  & \multirow{1}{*}{Description}                                                                                                & \multirow{1}{*} {Usage}                                                                        & \multirow{1}{*} {Literature}  \\
			\midrule
			Accuracy    &  \makecell[l]{Model accuracy on erased datasets and \\remaining datasets }                                                               & \makecell[l]{To evaluate the predictive accuracy \\ of the unlearned model  }                      &  \cite{bourtoule2021machine,guo2019certified}, ...                   \\
			\cline{1-4}
			Running time & The training time of unlearning process                                                                                                                      & To evaluate the unlearning efficiency                                                           & \cite{cao2015towards,guo2019certified}, ...     \\
			\cline{1-4}
			L2-norm      & \makecell[l]{ The parameters differences between the \\ retrained and unlearned models  $|| \theta_{1} - \theta_{2}||$  }                                      & \makecell[l]{To evaluate the indistinguishability \\ between two models}            & \cite{wu2020deltagrad}     \\
			\cline{1-4}
			KL-Divergence & \makecell[l]{ The KLD between the distribution of the \\ unlearned and retrained model: \\ $ \text{KL}( \mathcal{A}(D_r)||\mathcal{U}(D_e,  \mathcal{A}(D)))$ }          & \makecell[l]{To evaluate the indistinguishability  \\ between model parameters}            & \cite{nguyen2020variational}    \\
			\cline{1-4}
			JS-Divergence & \makecell[l]{ The distance between the  predictions \\ of retrained and unlearned model: \\ $JS(\mathcal{A}, \mathcal{U}) = 0.5 \cdot KL(\mathcal{A}||Q) + 0.5 \cdot KL(\mathcal{U}||Q)$ } & \makecell[l]{To evaluate the indistinguishability \\ between model outputs} & \cite{chundawat2022can}    \\
			\cline{1-4}
			Membership inference &  Recall ($\#$detected objects $/  \#$erased objects)                                                                                                     & \makecell[l]{ To verify if the erased sample is \\ unlearned by the model}          & \cite{graves2021amnesiac}    \\
			\cline{1-4}
			Epistemic uncertainty & efficacy$(\theta;D)= \left\{\begin{aligned} & \frac{1}{tr(I (\theta;D))}, \text{if } tr(I (\theta;D))>0  \\ &\infty, \text{otherwise} \end{aligned} \right.$   & \makecell[l]{To evaluate how much information \\ the model exposes } & \cite{becker2022evaluating}     \\
			\cline{1-4}
			EMA & \makecell[l]{ Ensembled Membership Auditing: Ensemble \\ multiple membership metrics and  utilizes  \\ Kolmogorov-Smirnov (KS)  statistical tools to \\obtain a final auditing score}  & \makecell[l]{ To verify if an unlearned model \\ memorizes a query dataset} & \cite{liu2020have}   \\
			\cline{1-4}
			MIB & \makecell[l]{ Membership Inference via Backdooring:  \\ Achieve membership inference for the  \\  backdoored data by querying a certain  \\ number of black-box queries}  & \makecell[l]{ To verify if the model unlearns the \\ backdoored data }       & \cite{hu2022membership}     \\
			\cline{1-4}
			Forgetting rate (FR)  & \makecell[l]{ $FR = \frac{AF-BF}{BT}$,  where AF, BF and BT are \\ defined below}                                                                    & \makecell[l]{ To measure the rate of samples that \\ are modified from member to \\ non-member after unlearning} & \cite{ma2022learn}                                                   \\
			\bottomrule
	\end{tabular}}
	\vspace{-6mm}
	\begin{tabbing}
	\scriptsize
	{$tr(I (\theta;D))$ is the trace of $I (\theta;D)$, and $I (\theta;D)$ is the Fisher Information matrix \cite{becker2022evaluating}.}
	\end{tabbing}
\end{table*}

\subsection{The Unlearning Verification Methods}

In Section \ref{pr}, we have introduced the L2-norm \cite{wu2020deltagrad}, KLD \cite{nguyen2020variational}, and privacy leakage as the unlearning verification metrics. 
The common metrics also include accuracy in assessing the performance of the unlearned model and running time in evaluating the unlearning efficiency. We have listed these evaluation metrics in Table \ref{tab_unl_verify}. Here, we will introduce some new evaluation metrics that recent studies tailored to unlearning. 


\medskip
\noindent
\textbf{Attack-based verification.}
Inspired by backdoor attacks in ML, \citeauthor{hu2022membership} \cite{hu2022membership} proposed Membership Inference via Backdooring (MIB).
MIB leverages the property of backdoor attacks that backdoor triggers will misadvise the trained model to predict the backdoored sample to other wrong classes. The main idea of MIB is that the user proactively adds the trigger to her data when publishing them online so that she can implement the backdoor attacks to determine if the model has been trained using her dataset.
MIB evaluates the membership inference for the triggered data by calculating the results of a certain number of black-box queries to the targetted model. Although MIB is effective to verify the unlearning of backdoored samples, it is hard to directly verify the unlearning of benign samples.


A similar membership-inference-based method was proposed in \cite{ma2022learn}. \citeauthor{ma2022learn}~\cite{ma2022learn} verified the effectiveness of unlearning methods by the proposed forgetting rate (FR) metric.
The evaluation metrics were defined using the observation of membership inference.
Suppose $D_e$ is the erased dataset; the FR of an unlearning method is denoted as $FR = \frac{AF-BF}{BT}$. In their FR definition, $BF$ and $AF$ are the samples in $D_e$ that are predicted as false by a membership inference attack before and after machine unlearning operations.
$BT$ is the sample size in $D_e$, which is indicated to be correct by a membership inference attack before machine unlearning.
According to the definition, $FR$ presents an instinctive evaluation of how many data points are altered from member to non-member by unlearning. 
If an unlearning method achieves that $AF > BF$ on the condition that $BT >0$, it means this method is effective. 
By contrast, the unlearning will be meaningless. The membership inference is an important verification tool, but failing an MIA is a necessary but not sufficient condition for successful unlearning.



\citeauthor{sommer2022athena} \cite{sommer2022athena} introduced ``Athena'', which leverages the property of backdooring techniques to verify the effectiveness of unlearning. Athena effectively and confidently certifies whether the data is deleted from an unlearning method. Thus, it provides a basis for quantitatively inferring unlearning. 
In their backdoor-based verification scheme, they first backdoor users' data and then test the backdoor success probability to infer if the data is unlearned. Like the MIB method, Athena is also hard to directly verify the unlearning of benign data, and the backdooring technique should be processed before the original model training.


\medskip
\noindent
\textbf{Model-centric (influence or model-difference) audits.} 
Another line of research views the difference between the model before and after unlearning as a meaningful source of evidence. EMU trains reconstruction models over simulated model differences to infer attributes of the deleted data, thereby measuring residual privacy leakage without depending on explicit backdoor design \citep{wang2025evaluation}. Building on this perspective, TAPE focuses on tailored posterior differences and enhances auditing with data perturbation and influence-based partitioning, enabling a more fine-grained assessment of how much private information about forgotten samples can still be recovered across different unlearning settings \citep{wang2025tape}. Similarly, TruVRF adopts a non-invasive, model-centric perspective by leveraging model sensitivity to verify unlearning at class-, volume-, and sample-level granularities, extending model-change-based auditing toward finer-grained white-box verification without relying on explicit backdoor design \citep{zhou2025truvrf}.
These approaches move auditing beyond direct output probing and toward more principled evidence grounded in influence patterns and model-change signatures.


\begin{table*}[t]
	\caption{The Employed Datasets in Machine Unlearning \vspace{-2mm}}
	\label{tab_datasets}
	\resizebox{\linewidth}{!}{
		\begin{tabular}{c|ccccc}
			\toprule
			\multirow{1}{*} {Data Type} & \multirow{1}{*}{Name of Datasets} &\multirow{1}{*}{Feature Dimension } & \multirow{1}{*} {\#. Samples} &\multirow{1}{*}{Task Type}& \multirow{1}{*}{Employed by}\\
			\midrule
			\multirow{7}{*} {Image} & MNIST~\cite{lecun1998gradient} & $28 \times 28 \times 1$ 		& $70,000$ & Classification & \cite{gupta2021adaptive,wang2023machine}, ...\\
			& CIFAR10    & $32 \times 32 \times 3$	 & $60,000$ & Classification & \cite{wang2023machine,tarun2023fast}, ...\\
			& CIFAR100  & $32 \times 32 \times 3$ 	& $60,000$ & Classification &\cite{tarun2023fast,cha2024learning}, ...\\
			& SVHN        & $32 \times 32 \times 3$ 		& $99,289$ & Classification &\cite{bourtoule2021machine,chundawat2023zero}, ...\\
			& ImageNet~\cite{deng2009imagenet} &  $224 \times 224 \times 3$ & $1,281,167$    & Classification &\cite{cha2024learning,tarun2023fast}, ...\\
			& GTSRB~\cite{stallkamp2011german}  &  $32 \times 32 \times 3$ & $51,839$    & Classification &\cite{wu2022federated,wei2023shared}, ...\\
			& Market-1501~\cite{zheng2015scalable} &  $128 \times 64 \times 3$ & $32,668$    & Person re-identification &\cite{mehta2022deep}\\
			\midrule
			\multirow{7}{*} {Tabular}   & Adult 	&  $14$ 										&  $48,842$   & Classification &\cite{chen2021machine,warnecke2024machine}, ...\\
			& Credit info &  $30$ 									& $284,807$   & Classification and anomaly detection&\cite{warnecke2024machine,du2019lifelong}, ...\\
			& Covtype 				&  $54$									& $581,012$  & Classification &\cite{mahadevan2021certifiable,wu2023deltaboost}, ...\\
			& HIGGS~\cite{baldi2014searching}  &  $28$									& $11,000,000$  &Binary classification &\cite{mahadevan2021certifiable,mahadevan2022certifiable}, ...\\
			& YELP2018 				&  $5$									& $1,561,406$  &Recommendation&\cite{chen2022recommendation}\\
			& Movielens-1m 				&  $4$									& $1,000,209$  &Recommendation&\cite{chen2022recommendation}\\
			& Movielens-10m 				&  $4$									& $10,000,054$  &Recommendation&\cite{chen2022recommendation}\\
			\midrule
			\multirow{2}{*} {Text}  & AG News~\cite{zhang2015character}	&  $3$									& $127,600$  & Text classification &\cite{wu2020deltagrad}\\
			& RCV1~\cite{glewis2004new} &  $3$									& $804,414$  & Text classification &\cite{wu2020deltagrad}\\
			\midrule
			\multirow{16}{*} {Graph}  & Amazon Photo~\cite{shchur2018pitfalls} & \makecell[c]{Features per Node: $745$}& \makecell[c]{Nodes: $7,650$ \\ Edges: $119,081$}& \makecell[c]{Node classification \\ Link prediction}  &\cite{chien2022certified,chien2022efficient}, ...\\
			\cmidrule(r){2-6}
			& Cora~\cite{sen2008collective} & \makecell[c]{Features per Node: $1,433$}& \makecell[c]{Nodes: $2,708$ \\ Edges: $5,429$}  & \makecell[c]{Node classification \\ Link prediction}   &\cite{chien2022efficient,chen2022graph}, ...\\
			\cmidrule(r){2-6}
			& Citseer~\cite{yang2016revisiting} & \makecell[c]{Features per Node: $3,703$}& \makecell[c]{Nodes: $3,327$ \\ Edges: $4,732$}  & \makecell[c]{Node classification \\ Link prediction}   &\cite{chen2022graph}\\
			\cmidrule(r){2-6}
			& Pubmed~\cite{fey2019fast} & \makecell[c]{Features per Node: $500$}& \makecell[c]{Nodes: $19,717$ \\ Edges: $44,338$}  & \makecell[c]{Node classification \\ Link prediction}   &\cite{chen2022graph,chien2022certified}, ...\\
			\cmidrule(r){2-6}
			& ogbn-arxiv~\cite{hu2020open} & \makecell[c]{Features per Node: $128$}& \makecell[c]{Nodes: $169,343$ \\ Edges: $1,166,243$}  & \makecell[c]{Node classification \\ Link prediction}   &\cite{chien2022efficient,chien2022certified}, ...\\
			\cmidrule(r){2-6}
			& Computers~\cite{mcauley2015image} & \makecell[c]{Features per Node: $767$}& \makecell[c]{Nodes: $13,752$ \\ Edges: $245,861$}  & \makecell[c]{Node classification \\ Link prediction}   &\cite{chen2022graph,chien2022certified}, ...\\
			\cmidrule(r){2-6}
			& CS~\cite{shchur2018pitfalls} & \makecell[c]{Features per Node: $767$}& \makecell[c]{Nodes: $18,333$ \\ Edges: $327,476$}  & \makecell[c]{Node classification \\ Link prediction}   &\cite{chen2022graph}\\
			\cmidrule(r){2-6}
			& Physics~\cite{shchur2018pitfalls} & \makecell[c]{Features per Node: Text}& \makecell[c]{Nodes: $27,770$ \\ Edges: $352,807$}  & \makecell[c]{Node classification \\ Link prediction}   &\cite{chen2022graph}\\
			\bottomrule
	\end{tabular}}
\end{table*}


\subsection{The Employed Datasets}

We collect the commonly employed datasets in machine unlearning studies and present the detail introduction of them in \Cref{tab_datasets}. There are four main types of data: Image, Tabular, Text and Graph. Most of the unlearning studies use image datasets and train classification models based on these image datasets. For tabular datasets, most of them are used in recommendation systems. The unlearning studies that investigate how to unlearn a recommendation model will use these tabular datasets. Graph data is employed for node classification and link prediction tasks, which is usually used in graph unlearning studies. For convenience to find the related studies, we link the corresponding unlearning studies at the last column in \Cref{tab_datasets}.

\section{Domain-Centric Machine Unlearning} \label{app_and_domain}

After introducing the classic centralized machine unlearning techniques and corresponding evaluation methods, in this section, we will introduce the studies of domain-centric unlearning, such as federated unlearning, graph unlearning, diffusion model unlearning, and large language model unlearning.

\subsection{Federated Unlearning} \label{new}


 \begin{figure}[h]
 	\centering
 	\includegraphics[width=0.95\linewidth]{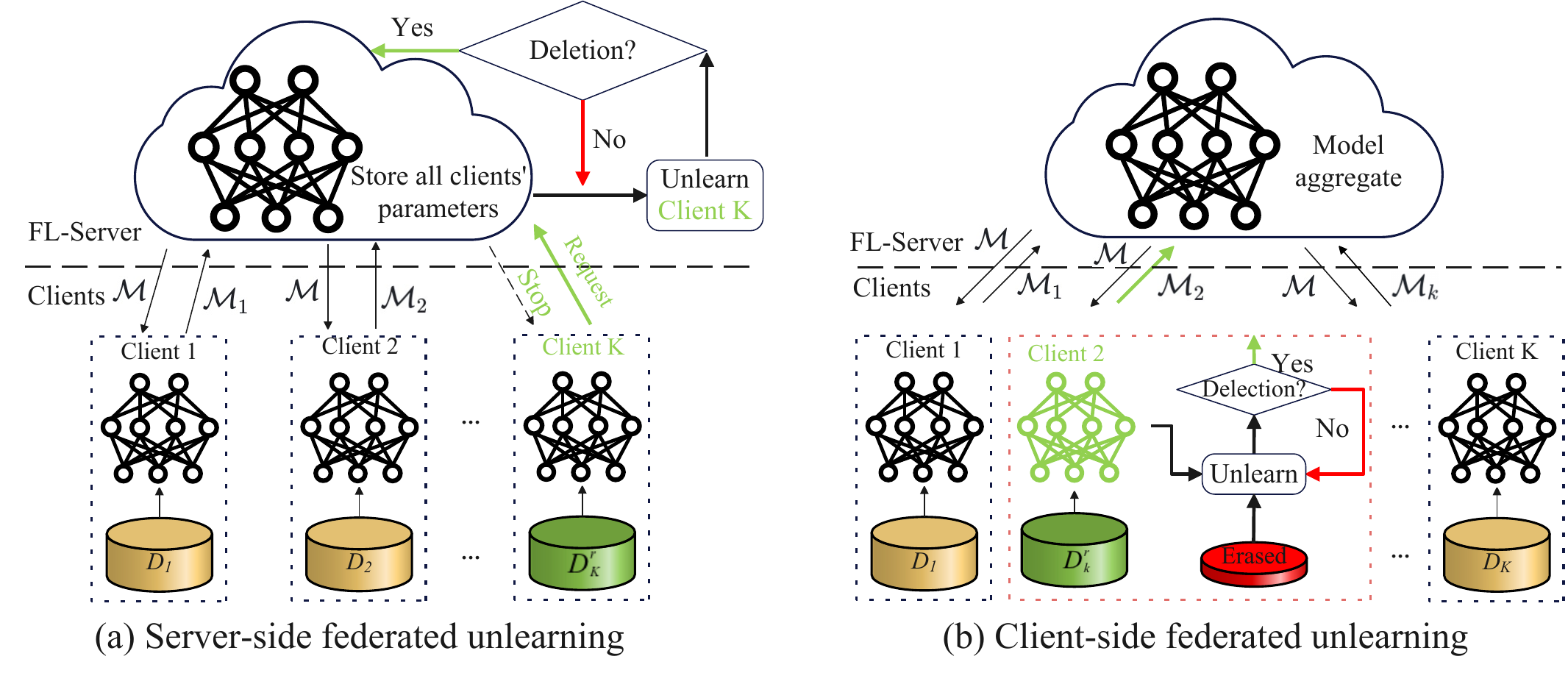}
 	\caption{Comparison between (a) server-side federated unlearning and (b) client-side federated unlearning}
 	\label{fig:fedunl}
 \end{figure}

FL was initially introduced to protect the privacy of participating clients during the machine learning training process in distributed settings. All participants will only upload their locally trained model parameters instead of their sensitive local data to the FL server during model training processes \citep{McMahanMRA16}. Therefore, in a federated learning scenario, limited access to the dataset will become a unique challenge when implementing unlearning. According to the unlearning target of a whole client's contribution or samples' contribution, we can roughly divide existing unlearning studies into two categories: client-level and sample-level federated unlearning. Since the client-level unlearning is usually operated in the server side and the sample-level unlearning usually needs the clients' participation, we can also call them server-side and client-side federated unlearning, as shown in \Cref{fig:fedunl}.

\noindent
\textbf{Client-level (Server-side) Federated Unlearning.}
Since the local data cannot be uploaded to the federated learning (FL) server side, most federated unlearning methods try to erase a certain client's contribution from the trained model by storing and estimating the contribution of uploaded parameters. In this situation, they can implement federated unlearning without interacting with the client, shown as the server-side federated unlearning in Fig. \ref{fig:fedunl} (a). The two representative methods are \cite{liu2021federaser, wu2022federated}.
\citeauthor{liu2021federaser} \cite{liu2021federaser} proposed ``FedEraser'' to sanitize the impact of a FL client on the global FL model. 
In particular, during FL training process, the FL-Server maintains the updates of the clients at each routine iteration and the index of the related round to calibrate the retrained updates. Based on these operations, they reconstructed the unlearned FL model instead of retraining a new model from scratch. However, FedEraser can only unlearn one client's data, which means it must unlearn all the contributions of this specific client's data. It is unsuitable for a client who wants to unlearn a small piece of his data. Study \cite{wu2022federated} tried to erase a client's influence from the FL model by removing the historical updates from the global model. 
They implemented federated unlearning by using knowledge distillation to restore the contribution of clients' models, which does not need to rely on clients' participation and any data restriction.

\noindent
\textbf{Smaple-level (Client-side) Federated Unlearning.}
Different from unlearning a whole client's influence and unlearning a class, \citeauthor{liu2022right} \cite{liu2022right,wang2023bfu} investigated how to unlearn data samples in FL, shown as the client-side federated unlearning in Fig. \ref{fig:fedunl} (b). In \cite{liu2022right}, they first defined a federated unlearning problem and proposed a fast retraining method to withdraw the influence of data from the FL model. Then, they proposed an efficient federated unlearning method following the Quasi-Newton methods and the first-order Taylor approximate method \cite{liu2022right}. 
They utilized the practical Fisher Information Matrix to model the Hessian matrix at a low cost. {Another similar work based on influence function to implement federated unlearning was introduced in \citep{wang2024fedu}.} Moreover, in \cite{wang2023bfu}, the authors implement federated unlearning based on Bayesian inference, and they propose a parameters self-sharing method to reduce the model utility degradation. {Although these methods effectively implement federated unlearning, they still need some benign clients to participate in the unlearning training, which limits the realistic deployment.}


\begin{figure}[h]
	\centering
	\includegraphics[width=0.98\linewidth]{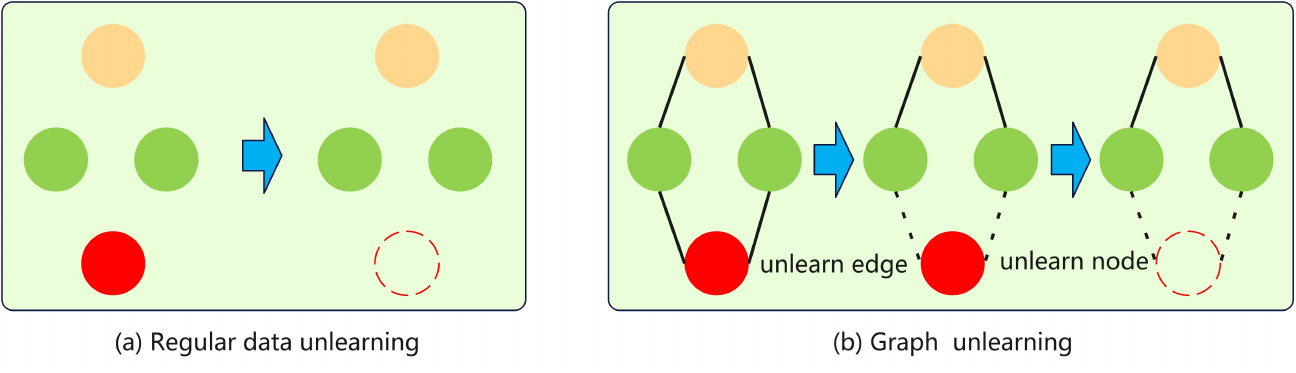}
	\caption{(a) Regular data unlearning, which only unlearns the data sample. (b) Graph unlearning. Since the graph data contains both node features and connective edge information, we may need to unlearn edge information or both edges and nodes in graph unlearning.}
	\label{fig:graphunl}
\end{figure}

\subsubsection{Graph Unlearning}
We introduce graph unlearning as a representative kind of irregular data unlearning. 
In \cite{chien2022certified,chen2022graph,cong2022privacy,zhang2024forgetting}, researchers extend regular data machine unlearning to a graph data scenario.
Graph structure data are more complex than standard structured data because graph data include not only the feature information of nodes but also the connectivity information of different nodes, shown in Fig. \ref{fig:graphunl}. 
Therefore, \citeauthor{chien2022certified} \cite{chien2022certified} proposed node unlearning, edge unlearning, and both node and edge unlearning for simple graph convolutions (SGC). Besides the different information unlearned in a graph learning problem, they found another challenge associated with feature mixing during propagation, which needs to be addressed to establish provable performance guarantees. 
They gave the theoretical analysis for certified unlearning of GNNs by illustrating the underlying investigation on their generalized PageRank (GPR) extensions and the example of SGC.

\citeauthor{chen2022graph} \cite{chen2022graph} found that applying SISA \cite{bourtoule2021machine} unlearning methods to graph data learning will severely harm the graph-structured information, resulting in model utility degradation.
Therefore, they proposed a method called GraphEraser to implement unlearning tailored to graph data.
Similar to SISA, they first cut off some connecting edges to split the total graph into some sub-graphs. Then, they trained the constituent graph models on these sub-graphs and ensembled them for the final prediction task. To realize graph unlearning efficiently, they proposed two graph partition algorithms and corresponding aggregation methods based on them.

In \cite{cong2022privacy}, \citeauthor{cong2022privacy} filled in the gap between regularly structured data unlearning and graph data unlearning by studying the unlearning problem on the linear-GNN. To remove the knowledge of a specified node, they design a projection-based unlearning approach, PROJECTOR, that projects the weights of the pre-trained model onto a subspace irrelevant to the deleted node features. PROJECTOR could overcome the challenges caused by node dependency and is guaranteed to unlearn the deleted node features from the pre-trained model.

\subsection{Unlearning Diffusion Models}

Exploring unlearning solutions for diffusion models and large language models (LLMs) is popular in recent years. We will first introduce the literature about unlearning methods for diffusion models and then introduce the LLMs unlearning in the next subsection.

Unlike the classic classification unlearning setting, unlearning for generative models must consider about open-ended conditional output distributions. The unlearned model should cease to reproduce specific training examples and protected artistic styles. Commonly, diffusion model unlearning methods can be organized by what is removed. Firstly, data-centric deletion targets the influence of designated training points and aims to approximate retraining on a pruned dataset~\citep{li2024machine,alberti2025data,chen2024score}. Secondly, concept-centric deletion removes higher-level semantic notions (e.g., nudity or protected styles) by intervening in the model's internal representations and denoising dynamics~\citep{wu2025erasing, zhang2024defensive, wu2025unlearning, cywinski2025saeuron}.

\medskip
\noindent
\textbf{Data-centric deletion.} 
A line of the diffusion model unlearning treats unlearning as an approximate data removal problem. The goal of these studies is to match the behaviour of a model retrained from scratch based only on the remaining dataset. Li et al. \citep{li2024machine} studied machine unlearning for image-to-image generative models. They contributed a general unlearning framework for structured translation models such as diffusion models, VQ-GAN, and MAE. They showed on ImageNet-1k and Places-365 that their method can effectively forget target samples with only a small degradation on retrained ones. Alberti et al. \citep{alberti2025data} focused on data unlearning in diffusion models. They defined retraining without the target data as the gold standard, then proposed SISS (Subtracted Importance Sampled Scores), a method that combines quality preservation and forgetting strength through importance sampling. Experiments on CelebA-HQ, MNIST with T-shirt, and Stable Diffusion show that SISS achieves a strong trade-off between forgetting and generation quality. Chen et al. \citep{chen2024score} proposed Score Forgetting Distillation (SFD), a data-free unlearning method for diffusion models. Instead of using original training data, SFD adds an unlearning objective into score distillation, encouraging the model to align the scores of unsafe classes.

\medskip
\noindent
\textbf{Concept-centric deletion.} 
A second line of diffusion unlearning studies targets semantic removal rather than individual data points. They usually aim to erase concepts or styles that may be harmful. EraseDiff \citep{wu2025erasing} formulates concept removal as a constrained optimization problem: the model should move away from denoising trajectories associated with the forget set, while staying close to behaviors needed for retained content. AdvUnlearn \citep{zhang2024defensive} shifts the focus to robustness, integrating adversarial training into the unlearning process so that erased concepts cannot be easily recovered through adversarial prompts. Importantly, it shows that editing the text encoder can yield stronger and more transferable robustness than editing the UNet alone. DoCo \citep{wu2025unlearning} further targets generalization and utility preservation by aligning the output domains of sensitive and anchor concepts and using gradient surgery to reduce interference with related benign concepts. 

In contrast to these parameter-level approaches, SAeUron \citep{cywinski2025saeuron} offers a more interpretable feature-level intervention by training sparse autoencoders on internal activations and blocking the features associated with unwanted concepts. Thereby, it enables transparent and multi-concept unlearning with competitive robustness and generation quality.

\medskip
\noindent
\textbf{Cross-cutting evaluation.} 
For diffusion models, unlearning evaluation should go beyond checking whether a target concept disappears in a few post-edit samples. A convincing evaluation framework must assess at least four properties: forgetting efficacy, utility preservation, robustness, and durability. Zhang et al.\ \citep{zhang2024unlearncanvas} make this point explicit by introducing UnlearnCanvas, a high-resolution benchmark for object and style unlearning that provides a standardised and automated evaluation pipeline with seven quantitative metrics. They also examine harder settings such as adversarial prompts, finer-grained concept removal, and sequential unlearning, showing that immediate deletion performance alone is insufficient for judging success. Ko et al.\ \citep{ko2024boosting} further show that diffusion unlearning can substantially impair text--image alignment, and therefore argue that evaluation must jointly measure forgetting and alignment rather than treating utility as generic sample quality alone. Their post-unlearning alignment framework is motivated precisely by this tension. George et al.\ \citep{george2025illusion} extend evaluation to the temporal dimension by showing that supposedly erased concepts can re-emerge after benign fine-tuning.


These studies suggest that diffusion unlearning should be evaluated not only by how much is forgotten immediately after editing, but also by whether the model retains non-target capabilities, resists adversarial recovery, and remains stably unlearned after subsequent updates.

\subsection{Unlearning Large Language Models}

Unlearning in large language models (LLMs) seeks to remove the concepts and behaviours, such as copyrighted text, personal information, and unsafe responses. Recent studies conceptualize this problem as eliminating undesirable training influence and the capabilities associated with it under two key requirements: locality and utility, meaning that changes should be confined to the targeted knowledge and general performance on unrelated tasks should remain largely intact~\citep{yao2024large,yao2024machine,liu2025rethinking}. From this perspective, existing approaches can be broadly divided into two categories according to where forgetting is implemented. The first is parametric post-hoc unlearning, which directly updates model parameters so that forgetting becomes embedded within the model itself  \citep{yao2024machine, yao2024large, liu2024towards, huang2024offset, wang2025selective}. The second is non-parametric or suppression-based unlearning, which leaves the underlying model unchanged and instead prevents undesirable outputs through inference-time intervention or control \citep{liu2024large}.

\noindent
\textbf{Parametric post-hoc unlearning.}
Most LLMs unlearning studies are about post-hoc unlearning directly on pretrained LLMs by updating weights to suppress targeted information. Large Language Model Unlearning \citep{yao2024large} presents unlearning primarily as an alignment mechanism, arguing that targeted forgetting can suppress harmful outputs, remove copyrighted content, and reduce hallucinations using only negative examples. Machine Unlearning of Pre-trained Large Language Models \citep{yao2024machine}, in contrast, takes a more systematic and benchmark-driven view, evaluating multiple unlearning strategies on pre-trained models and showing that gradient-based post-hoc unlearning can be highly efficient relative to retraining, especially when forgetting updates are combined with utility-preserving regularisation on retained data.

Building on early gradient-based approaches, subsequent work develops more targeted unlearning mechanisms to better satisfy the locality requirement. Liu et al. \citep{liu2024towards} propose Selective Knowledge-negation Unlearning (SKU), a two-stage framework that first identifies harmful knowledge and then explicitly negates it, aiming to suppress harmful responses while preserving normal utility. Wang et al. \citep{wang2025selective} further refine the granularity of forgetting with SeUL, which performs unlearning at the level of sensitive spans instead of whole sequences and introduces dedicated metrics to quantify sensitive-information deletion. These methods reflect a broader shift from coarse unlearning toward more selective and locality-aware interventions, although they differ in whether forgetting is enforced through parameter negation, output-space offsets, or span-level objective design.

\medskip
\noindent
\textbf{Non-parametric unlearning (inference-time control).} 
Unlike parametric unlearning methods that edit model weights, Liu et al. \citep{liu2024large} shift forgetting to the inference stage. Their Embedding-Corrupted (ECO) framework treats unlearning as a conditional control problem: the base model remains intact, while a detector identifies prompts associated with forgotten content and an embedding-level intervention steers generation away from that content. In this sense, ECO does not make forgetting intrinsic to the model; rather, it imposes an “unlearned state” on demand for selected inputs. This design is attractive operationally because it is modular, scalable, and easy to deploy across different LLM backbones, especially in settings where direct access to model parameters is unavailable or repeated fine-tuning would be too costly.

\medskip
\noindent
\textbf{Evaluation infrastructure for LLMs unlearning.} 
Across recent work, evaluation in LLM unlearning is becoming both broader and more demanding. Jin et al. \citep{jin2024rwku} contribute a dedicated text-only benchmark, RWKU, which evaluates forgetting of real-world entity knowledge on LLaMA3-Instruct (8B) and Phi-3 Mini-4K-Instruct (3.8B) using membership-inference attacks, adversarial probes, and retain tasks drawn from MMLU, BBH, TruthfulQA, TriviaQA, and AlpacaEval. Liu et al. \citep{liu2025rethinking} provide a more general evaluation framework, arguing that unlearning should be assessed not only by immediate deletion success but also by scope precision, privacy leakage, robustness, consistency, and effects on causally unrelated capabilities. Extending this agenda beyond text-only models, Li et al. \citep{li2024single} propose SIU for multimodal LLMs and introduce MMUBench, built from concepts sampled from MIKE, image collections from web search, and evaluation on LLaVA 7B/13B together with standard multimodal benchmarks such as GQA, VQA-v2, VizWiz, ScienceQA-IMG, TextVQA, POPE, MMBench, and MM-Vet. Taken together, these studies show that unlearning evaluation is moving from coarse deletion scores toward more rigorous, attack-aware, locality-sensitive, and increasingly multimodal assessment.

\section{Privacy and Security Issues on Machine Unlearning} \label{priacy}

Although unlearning methods were initially proposed to safeguard users' privacy,  many researchers have noticed that unlearning brings new privacy threats simultaneously. In this section, we will discuss these privacy and security studies about machine unlearning.


\subsection{Privacy Threats on Machine Unlearning}

\subsubsection{Membership Inference Attacks in Unlearning}

\citeauthor{chen2021machine} \cite{chen2021machine} first pointed out that when a model is unlearned, the discrepancy in the outputs from the model before and after unlearning leaks privacy of erased data. 
Then, they proposed the corresponding membership inference attack pipeline in unlearning, which includes three phases: posterior generation, feature construction, and membership inference.
\begin{enumerate}
	\item  Posteriors Generation. Suppose that the attacker can access two versions of the trained model, the model $\theta_{\mathcal{A}}$ before unlearning and the model $\theta_{\mathcal{U}}$ after unlearning. Assume a target data point $e$, the attacker queries $\theta_{\mathcal{A}}$ and $\theta_{\mathcal{U}}$, and has the corresponding posteriors, $p(\theta_{\mathcal{A}})$ and $p(\theta_{\mathcal{U}})$, which also called as confidence values in \cite{shokri2017membership}.
	\item  Feature Construction. After achieving the two posteriors $p(\theta_{\mathcal{A}})$ and $p(\theta_{\mathcal{U}})$, the attacker sums them to make the inference feature vector $F$.  Common methods exist to construct the feature vector shown in \cite{chen2021machine}.
	\item  Inference. After the attacker finishes training the attack model based on the created features $F$, he inputs the collected feature to the inference model to predict if the specific sample $e$ occurs in the erased dataset of unlearning models. 
\end{enumerate}

In \cite{chen2021machine}, they assumed that the attacker has admission to two ML models before and after unlearning, but it is sometimes impractical, especially in black-box learning scenarios. \citeauthor{lu2022label} \cite{lu2022label} further proposed a label-only membership inference method to imply if a sample is unlearned, eliminating the dependence on accessing posteriors. 
Their basic idea is that the same noise injection on candidate data points will show different results for the sample in or not in the training dataset.
Thus, they made the adversary continuously query the original and unlearned models and add noise to modify their outputs. Observing the disturbance amplitude lets them determine whether an item is deleted.


\citeauthor{golatkar2020forgetting} \cite{golatkar2020forgetting} derived an upper bound to confirm the maximizing knowledge that can be extracted from a black-box model. They queried the model with a picture and obtained the related output. They used the entropy of the result probabilities to construct an effective black-box membership inference \cite{shokri2015privacy} attack in machine unlearning.

\subsubsection{Privacy Reconstruction Attacks in Unlearning}
Privacy reconstruction is another popular attack in machine unlearning. In an unlearning scenario, \citeauthor{gao2022deletion}~\cite{gao2022deletion} proposed the deleted reconstruction attacks to recover the removed data from the outputs of the original and unlearning models. 
In their work, they formalized erasure inference and erasure recovering attacks. The attacker seeks to infer which sample is removed or recovers the erased sample. 
In particular, for the deletion inference, they formulate the objective of an erasure inference to decide if a data instance $e$ was in or not in the erased dataset, $e \in D_e$ or $e' \notin D_e$. 
For the deletion reconstruction, they focused on reconstructing the erased example $e$. In all their reconstruction attacks, the attacker does not have any particular samples, and the purpose is to extract the features knowledge of the erased example. Specifically, the deletion reconstructions include the deleted instance reconstruction and deleted label reconstruction. As named, the deleted instance reconstruction is to extract all of the information of the erased example, and the erased label reconstruction is to infer the label of the erased point in the classification problem.

\citeauthor{zanella2020analyzing} \cite{zanella2020analyzing} indicated that the releasing snapshot of overlapped language models would leak the privacy of the training dataset. They verified that the model updates significantly threaten the private information added to or deleted from the training dataset by many experimental results.
\citeauthor{zanella2020analyzing} found five phenomena. First, an attacker can extract particular sentences used or not in the training dataset by comparing two models. Second, analyzing more model snapshots shows more information about the updated data than considering fewer model snapshots. 
Third, adding or deleting other non-private data during model updates can not mitigate privacy leakage. 
Fourth, differential privacy can reduce privacy leakage risks and decrease trained models' accuracy. 
Fifth, to mitigate the privacy leakage risks while keeping the model utility, the server can limit the model parameters access or only output a subset of the results.

Many studies further utilized these privacy threats to evaluate the unlearning effectiveness.
\citeauthor{huang2021mathsf} \cite{huang2021mathsf} proposed Ensembled Membership Auditing (EMA) for auditing data erasure. They use the membership inference to assess the removing effectiveness of unlearning.
\citeauthor{graves2021amnesiac} \cite{graves2021amnesiac} indicated that if an attacker can infer the sensitive information that was wanted to be erased, then it means that the server has not guarded the rights to be forgotten.
\citeauthor{baumhauer2022machine} \cite{baumhauer2022machine} developed linear filtration to sanitize classification models with logits prediction after class-wide deletion requests. 
They verified their methods by testing how well the method defends against privacy attacks.

Both \cite{chen2021machine} and \cite{zanella2020analyzing} pointed out that differential privacy guarantees that a model does not reveal too much knowledge about any training sample.
A DP-protected model can further guarantee the group's privacy by binding the impacts of a bunch of training samples.
If using DP to protect the privacy of a bunch of $|D \backslash D_e| $training examples against snapshot attacks on $\theta_D, \theta_{D\backslash D_e}$, it means that $\theta_{D \backslash D_e}$ cannot be more useful than $\theta_D$.




\subsection{Security Threats on Machine Unlearning}

There are several papers that introduced security threats targeting machine unlearning. For instance, \citeauthor{marchant2022hard}~\cite{marchant2022hard} examine an attacker's strategy to increase the computational cost of performing machine unlearning. Studies by \citeauthor{zhao2024static} \cite{zhao2024static} and \citeauthor{hu2023duty} \cite{hu2023duty} investigate security vulnerabilities within unlearning systems, specifically by uploading customized malicious data update requests to negatively influence the model utility. Additionally, some studies propose backdooring or poisoning methods that exploit unlearning requests to achieve backdoor insertion, as discussed by \cite{di2022hidden} and \cite{liu2024backdoor}. Huang et al. \citep{huang2024uba} introduced an unlearning-activated backdoor attack that uses influence-based analysis to select camouflage to trigger samples, so that the backdoor remains stealthy before unlearning but becomes effective after specific deletions have taken place. These works collectively highlight the need for robust security measures in machine unlearning to mitigate these emerging threats.

\subsection{Defending Methods}

From a defensive standpoint, only a limited number of studies have begun to design privacy-preserving and robust unlearning methods to defend against these emerging threats. Representative examples are BlindU \citep{wang2026blindu} and Compressive Representation Forgetting (CRFU) \citep{crfu25wwq}. CRFU mitigates reconstruction attacks and related forms of information leakage attacks by imposing an information-bottleneck-inspired compression objective on the representations of data designated for removal \citep{crfu25wwq}. In doing so, CRFU aims to suppress the mutual information between latent representations and the erased inputs, while preserving model utility through remembering objectives and a controllable unlearning rate \citep{crfu25wwq}. BlindU provides a stricter privacy-preserving unlearning framework compared with CRFU. Privacy preservation of BlindU is achieved by uploading compressed and differential privacy-enabled data as the unlearning requests, and an unlearning method in BlindU for compressed data is designed correspondingly \citep{wang2026blindu}. More broadly, this line of work signals a move away from viewing unlearning as a simple post hoc algorithmic adjustment, and toward framing it as a joint optimisation problem that simultaneously addresses forgetting, privacy protection, and robustness.

\section{Machine Unlearning Applications} \label{app}

\begin{table*}
	\caption{Machine Unlearning Application \vspace{-2mm}}
	\label{tab:applications}
	\resizebox{\linewidth}{!}{
		\begin{tabular}{cccccc}
			\toprule
			\multirow{1}{*} {Literature} & \multirow{1}{*}{Application Scenarios} &\multirow{1}{*}{Methods } & \multirow{1}{*} {Realization Methods} &\multirow{1}{*}{Evaluation Metric}& \multirow{1}{*}{Year}\\
			\midrule
			\cite{liu2022backdoor} & Mitigate backdoor& BAERASER&Gradient ascent method & ASR and ACC &2022\\
			\cite{wang2019neural} & Mitigate backdoor& -&Median Absolute Deviation &L1-Norm, $\#$FP, ACC, ASR &2019\\
			\cite{du2019lifelong} & Anomaly detection & - &Maintain memory set& $\#$FP and $\#$FN &2019\\
			\cite{cao2018efficient} & Repair pollution& KARMA& Cluster and unlearn & DA &2018\\
			\cite{huang2021unlearnable} & Data ``unlearnable'' & EMP &Add noise & ACC &2021\\
			\bottomrule
	\end{tabular}}
	\begin{tablenotes}
		\scriptsize
		\normalsize
		\item ASR:Attack Success Rate, ACC: Accuracy
		\item DA:Detection Accuracy, EMP: Error-minimizing Perturbations
		\item $\#$FN: number of false negative, $\#$FP: number of false positive
	\end{tablenotes}
\end{table*}

Besides the inherent demands of machine unlearning that draw much research attention, many researchers also find that machine unlearning can be applied in many other scenarios to solve related problems. We list the recent unlearning applications in Table \ref{tab:applications}. The most popular application of machine unlearning is to mitigate the anomalies, including backdoor triggers and pollution from an already-trained model.

\noindent
\textbf{Mitigating Anomaly.}
Study \cite{wang2019neural} applied machine unlearning in detecting and mitigating backdoor attacks in DNNs.
They designed two approaches to eliminate backdoored neurons from the backdoored model and repair the model to be strong against malicious pictures. 
Then, in \cite{liu2022backdoor}, they offered that a backdoor model learned the poisoned decision boundary. Data points with triggers are all classified into the target class.
They reversed the backdoor injection process to defend against it in machine unlearning, which is simple but effective. 
Their method contains two primary steps: first, they utilize a max-entropy staircase approximator to complete trigger reconstruction; second, they remove the added triggers using unlearning. They named these two key steps of BAERASER as trigger pattern recovery and trigger pattern unlearning. 
Via a dynamic penalty mechanism, they mitigated the sharp accuracy degradation of gradient-ascent-based machine unlearning methods.

Repairing pollution is another successful unlearning application. \citeauthor{cao2018efficient} \cite{cao2018efficient} proposed ``KARMA'' to search various subsets of original datasets and return the subset with the highest misclassifications. First, KARMA searches for possible reasons that lead to the wrong ML model classification. It clusters the misclassified samples into various domains and extracts the middle of clusters. KARMA prioritizes the search for matching examples in the original datasets using these extracted centers. Second, KARMA  grows the reason discovered in the first step by discovering more training samples and creating a cluster. Third, KARMA determines if a causality cluster is polluted and calculates how many samples the cluster contains. \citeauthor{du2019lifelong} \cite{du2019lifelong} also explored unlearning in lifelong anomaly detection and tried to mitigating exploding loss and sharp accuracy degradation caused by unlearning.


\noindent
\textbf{Data Unlearnable.}
\citeauthor{huang2021unlearnable} \cite{huang2021unlearnable} presented a method that can make samples unlearnable by injecting error-minimizing noise. This noise is intentionally synthesized to diminish the error of the samples close to zero, which can mislead the model into considering there is "nothing" to learn. They first tried the error-maximizing noise but found that it could not prevent DNN learning when used sample-wise to the training data points. Therefore, they then begin to study the opposite direction of error-maximizing noise. In particular, they proposed the error-minimizing noise to stop the model from being punished by the loss function during traditional ML model training. Therefore, it can mislead the ML model to consider that there is "nothing" to learn.

\section{Lessons Learnt and Discussions} \label{discuss}

In the research domain of machine unlearning, researchers mainly face three difficult challenges, the stochasticity of training, incrementality of training, and catastrophe of unlearning. Researchers tried many mechanisms to mitigate the influence of these challenges. For example, in exact unlearning, they designed split algorithms that divide the final model into several sub-models and avoid the stochasticity and incrementality of one sub-model to influence other sub-models \citep{bourtoule2021machine}. In approximate unlearning, they bounded the removed estimation to avoid accuracy degradation \citep{guo2019certified}. However, it is still not easy to analyze them clearly and solve them totally. As the research in machine unlearning in-depth, researchers extended it to new situations, such as federated learning and graph learning, and they met corresponding new challenges. Here, we discuss the differences between various unlearning scenarios and their corresponding challenges. Based on these challenges, we list some potential research directions on which we can focus.

In the common centralized unlearning scenario, retraining from scratch can achieve the best unlearning effect, but it is expensive in both computation and storage. Existing methods try to design new unlearning mechanisms to reduce the cost from the two aspects. {Although researchers proposed many methods, they just solved the challenges partially and cannot guarantee the final unlearning effect. Here, we find and list some open questions in traditional centralized unlearning.

\begin{enumerate}
	\item \textbf{Challenge:} The stochasticity of training and incrementality of training challenges are hard to solve exactly, so recent research mainly focuses on reducing utility degradation after unlearning. \textbf{Potential Direction:} Exploring the unlearning problem with a theoretical guarantee, such as from a differential geometry perspective, will provide the theoretical contribution and explanation to unlearning.
	\item  \textbf{Challenge:} What unlearning mechanisms should be to forget a certain or exact goal in different models, such as in diffusion model or LLMs?
	\textbf{Potential Direction:} Exploring to what extent we have to unlearn for different unlearning purposes remains under-explored, and how to utilize the learning mechanism to implement unlearning purposes in also an interesting direction.
\end{enumerate}
}


Machine unlearning in distributed scenarios has many differences from centralized scenarios. We take federated unlearning as a representative example of distributed unlearning.
The first difference is that federated unlearning can only be implemented locally on the client's side if they want to unlearn some specific samples because clients do not upload their data to the FL server in a federated scenario. To avoid interacting with clients during unlearning, researchers \cite{liu2021federaser, wu2022federated} proposed to unlearn the contribution of a whole client while not some samples of the client.
The second difference is that when unlearning requests come during the FL training process, the FL server must first execute the unlearning process and broadcast the unlearned model for later updating to avoid other clients wasting computation on the before-unlearned model. The third difference is that federated unlearning is more vulnerable to catastrophic degradation than centralized unlearning because if the FL server broadcasts the catastrophic unlearned model and other clients update based on the unlearned model, it will vanish the efforts of other clients that trained before. After introducing these differences, we can see that the challenges in federated unlearning are more complex than in centralized unlearning, and we conclude the following open problems in federated unlearning. 


\begin{enumerate}
	\item  \textbf{Challenge:} Federated unlearning is hard to be implemented by using the fast retraining methods because data is out of reach for the server. Most federated unlearning methods are implemented by using approximate unlearning methods. However, as we know, approximate unlearning easily causes catastrophic unlearning, and federated learning is more vulnerable to degradation. \textbf{Potential Direction:} Controlling the catastrophic in federated unlearning will be more urgent than in centralized unlearning.
	\item  \textbf{Challenge:} Existing federated unlearning methods require the activation of all users, including those without unlearning requests, to assist in the unlearning process. These approaches are impractical and inefficient, particularly when unlearning requests are frequent. \textbf{Potential Direction:} It is crucial to study how to balance the unlearning effectiveness and efficiency in federated unlearning. 
\end{enumerate}

Besides exploring machine unlearning based on regularly structured data, researchers tried to implement unlearning in graph data. Exact unlearning may be suitable for centralized graph unlearning if graph data is sparse. However, the challenges of approximate unlearning may be more difficult than structured-data-based unlearning because, in graph unlearning, the relationship and influence between data samples are more complex than structured data \cite{chen2022graph, chien2022certified}. In particular, graph data includes not only the node feature value but also the connecting edge information. Therefore, in graph unlearning, the estimation of the contribution of a node will be more difficult than in regular data unlearning. The original problems in regular data unlearning will be more challenging in graph unlearning. Besides these problems, graph unlearning also faces unique problems that are related to edge structure information. In graph unlearning, unlearning some edges or sub-graphs is a big question.

	Unlearning in LLMs and diffusion models introduces additional challenges beyond those in conventional discriminative models. In these generative models, the target to be removed is often distributed across parameters and entangled with related knowledge, so forgetting a specific sample, entity, or concept does not necessarily prevent the model from reproducing semantically related variants \citep{liu2025rethinking,chen2024score}. Existing evaluation is also still immature: LLM unlearning is heavily concentrated on a small number of fixed forget-retain benchmarks \citep{jin2024rwku}, while diffusion unlearning studies have shown that erased concepts may reappear under nearby or adversarial prompts \citep{george2025illusion}. Therefore, future research should focus on defining unlearning targets and scopes more precisely, such as at the entity, concept, and hierarchical knowledge levels, developing more realistic and robust benchmarks, improving continual and black-box unlearning for real deployment settings, and enhancing robustness and interpretability.

After designing machine unlearning algorithms, effective verification and auditing methods are necessary~\cite{thudi2022necessity}. Most existing unlearning verification methods rely on backdooring techniques. However, these methods inherently degrade model utility because they require mixing backdoored samples into the model training process. Investigating ways to preserve model utility in backdoor-based unlearning verification methods is an under-explored area. Additionally, backdoor-based verification methods must include backdoored samples in the unlearning requests to verify the effectiveness of these requests. This requirement restricts these methods from supporting verification for single-sample unlearning requests. Exploring new strategies that can effectively verify unlearning requests without compromising model utility or being suitable for both single-sample and multi-sample unlearning scenarios remains a significant challenge in this field.


Another important part is the privacy and security issues in machine unlearning. Machine unlearning was first proposed to protect users' privacy, but it brings new threats in that adversaries have a chance to infer the information about the removed data.
Literature \cite{chen2021machine,zanella2020analyzing} have pointed out that updates of unlearning will leak privacy information, and they proposed corresponding attacks to infer this private information. However, the most recent unlearning privacy leakage attacks have been similar to those in a learning situation. An attack that is tailored to unlearning mechanisms is expected. A similar situation exists in unlearning applications. Although researchers proposed to unlearn a backdoor trigger \cite{liu2022backdoor} or pollution \cite{cao2018efficient}, they only used a few unlearning techniques and paid more effort to detect those anomalies. One important reason is that the unlearning mechanism is not mature enough. We are at the beginning of machine unlearning investigation, and there are still many under-explored problems in unlearning itself. Finding machine unlearning applying situation and tailoring unlearning techniques to this situation is the direction of unlearning application.

\section{Summary} \label{conclu}
The survey aims to offer a comprehensive and systematic overview of machine unlearning techniques. We organize the main challenges, research advancements, corresponding techniques, and privacy and security issues in machine unlearning. Additionally, we presented a detailed and unified classification of machine unlearning. We first illustrate the complete unlearning framework, including the learning, request, unlearning, and verification. Then, we briefly categorize recent studies into exact and approximate unlearning and introduce the technical details of the corresponding unlearning methods. Moreover, we noticed some new unlearning scenarios, such as graph unlearning, federated unlearning, diffusion model unlearning, and LLMs unlearning, which are also introduced in the survey. Besides, we consider privacy and security issues in machine unlearning to be an important part of the studies. We collect and summarize the related literature about unlearning privacy threats and applications. Ultimately, the survey provides clear summaries and comparisons between various unlearning scenarios and corresponding methods, giving a comprehensive picture of existing work and listing the challenges and open problems of different scenarios of machine unlearning.

We hope our survey can help classify future unlearning studies, achieve a more in-depth understanding of unlearning methods, and address complex challenges. We believe the open problems listed in Section \ref{discuss} will still be challenging in the following years, and we will try to optimize some of them. Last but not least, we expect the survey can help researchers in the study of machine unlearning, regardless of the unlearning strategies or unlearning privacy and security threats or applications of unlearning.

 \bibliographystyle{unsrtnat} 
 \bibliography{acmart}

\end{document}